\documentclass[longauth]{aa} % for the long lists of affiliations 
\usepackage{graphicx}
\usepackage{txfonts,xcolor}
\usepackage{float}
\usepackage{booktabs}
\begin{document}

   \title{The GRAVITY  young stellar object survey }
   \subtitle{III. The dusty disk of RY Lup\thanks{GTO program with run ID 099.C-0667}}
\titlerunning{The dusty disk of RY~Lup}
 \authorrunning{Bouarour et al.}
 
  \author{GRAVITY Collaboration: Y.-I. Bouarour\inst{1,2,3}, K. Perraut\inst{1}, F. M\'enard\inst{1}, W. Brandner\inst{5}, A. Caratti o Garatti\inst{2,3}, P. Caselli\inst{6}, E. van Dishoeck\inst{6,15}, C. Dougados\inst{1}, R. Garcia-Lopez\inst{2,3,5}, R. Grellmann\inst{4}, T. Henning\inst{5}, L. Klarmann\inst{5}, L. Labadie\inst{4}, A. Natta\inst{2}, J. Sanchez-Bermudez\inst{12,5}, W.-F. Thi\inst{6}, P.T. de Zeeuw\inst{6,15} \and A. Amorim\inst{8,14} \and M. Bauböck\inst{6}, M. Benisty\inst{1} \and J.-P. Berger\inst{1} \and Y. Clenet\inst{11} \and V. Coudé du Foresto\inst{11} \and G. Duvert\inst{1} \and A. Eckart\inst{4,16} \and F. Eisenhauer\inst{6}, F. Eupen\inst{4} \and M. Filho\inst{8,9} \and F. Gao\inst{6} \and P. Garcia\inst{8,9} \and E. Gendron\inst{11} \and R. Genzel\inst{6} \and S. Gillessen\inst{6} \and A. Jiménez-Rosales\inst{6}, L. Jocou\inst{1} \and S. Hippler\inst{5} and M. Horrobin\inst{4}, Z. Hubert\inst{1} \and P. Kervella\inst{11} \and S. Lacour\inst{11} \and J.-B. Le Bouquin\inst{1} \and P. Léna\inst{11} \and T. Ott\inst{6} \and T. Paumard\inst{11} \and G. Perrin\inst{11} \and O. Pfuhl\inst{13}, G. Rousset\inst{11} \and S. Scheithauer\inst{5} J. Shangguan\inst{6}, J. Stadler\inst{6} \and O. Straub\inst{6} \and C. Straubmeier\inst{4} \and E. Sturm\inst{6} \and F. H. Vincent\inst{11} \and S.D. von Fellenberg\inst{6} \and F. Widmann\inst{6}, M. Wiest\inst{4}
}
\institute{Univ. Grenoble Alpes, CNRS, IPAG, F-38000 Grenoble, France\\  \email{youcef.bouarour@ucdconnect.ie}
     \and 
    Dublin Institute for Advanced Studies, 31 Fitzwilliam Place, D02\,XF86 Dublin, Ireland
    \and
    School of Physics, University College Dublin, Belfield, Dublin 4, Ireland. 
    \and
    I. Physikalisches Institut, Universit\"at zu K\"oln, Z\"ulpicher Strasse 77, 50937, K\"oln, Germany
\and
Max Planck Institute for Astronomy, K\"onigstuhl 17,
69117 Heidelberg, Germany
\and
Max Planck Institute for Extraterrestrial Physics, Giessenbachstrasse, 85741 Garching bei M\"{u}nchen, Germany
\and
Unidad Mixta Internacional Franco-Chilena de Astronom\'ia (CNRS UMI 3386), Departamento de Astronom\'ia, Universidad de Chile, Camino El Observatorio 1515, Las Condes, Santiago, Chile     
\and
CENTRA, Centro de Astrof\'{\i}sica e Gravita\c{c}\~{a}o, Instituto Superior T\'{e}cnico, Avenida Rovisco Pais 1, 1049 Lisboa, Portugal
\and
Universidade do Porto, Faculdade de Engenharia, Rua Dr. Roberto Frias, 4200-465 Porto, Portugal
\and 
European Southern Observatory, Casilla 19001, Santiago 19, Chile
\and
LESIA, Observatoire de Paris, Université PSL, CNRS, Sorbonne Université, Université de Paris, 5 Place Jules Janssen, 92195 Meudon, France
\and
Instituto de Astronom\'ia, Universidad Nacional Aut\'onoma de M\'exico, Apdo. Postal 70264, Ciudad de M\'exico 04510, Mexico
\and
European Southern Observatory, Karl-Schwarzschild-Str. 2, 85748
Garching, Germany
\and
Universidade de Lisboa - Faculdade de Ci\^encias, Campo Grande, 1749-016 Lisboa, Portugal
\and
Leiden Observatory, Leiden University, Postbus 9513, 2300 RA
Leiden, The Netherlands
\and
Max-Planck-Institute for Radio Astronomy, Auf dem Hügel 69, 53121, Bonn, Germany 
}

   \date{Received ; accepted }

% \abstract{}{}{}{}{} 
% 5 {} token are mandatory
\abstract
{Studies of the dust distribution, composition, and evolution of protoplanetary disks provide clues for understanding planet formation. However, little is known about the innermost regions of  disks where telluric planets are expected to form.}
{We aim constrain the geometry of the inner disk of the T~Tauri star RY~Lup by combining spectro-photometric data and interferometric observations in the near-infrared (NIR) collected at the Very Large Telescope Interferometer (VLTI). We use PIONIER data from the ESO archive and GRAVITY data that were obtained in June 2017 with the four 8m telescopes.}
{We use a parametric disk model and the 3D radiative transfer code MCFOST to reproduce the spectral energy distribution (SED) and match the interferometric observations. MCFOST produces synthetic SEDs and intensity maps at different wavelengths from which we compute the modeled interferometric visibilities and closure phases through Fourier transform.}
{To match the SED from the blue to the millimetric range, our model requires a stellar luminosity of  2.5~L$_{\odot}$,  higher than any previously determined values. Such a high value is needed to accommodate the circumstellar extinction caused by the highly inclined disk, which has been neglected in previous studies. While using an effective temperature of 4800~K determined through high-resolution spectroscopy, we derive a stellar radius of  2.29~R$_{\odot}$. These revised fundamental parameters, when combined with the mass estimates available (in the range 1.3-1.5~M$_{\odot}$), lead to an age of 0.5-2.0~Ma for RY~Lup, in better agreement with the age of the Lupus association than previous determinations. Our disk model (that has a transition disk geometry) nicely reproduces the interferometric GRAVITY data and is in good agreement with the PIONIER ones. We derive an inner rim location at 0.12~au from the central star.  This model corresponds to an inclination of the inner disk of 50$^\circ$, which is in mild tension with previous determinations of a more inclined outer disk from SPHERE (70$^\circ$ in NIR) and ALMA (67~$\pm$~5$^\circ$) images, but consistent with the inclination determination from the ALMA CO spectra (55~$\pm$~5$^\circ$). Increasing the inclination of the inner disk to 70$^\circ$ leads to a higher line-of-sight extinction and therefore requires a higher stellar luminosity of {4.65}~L$_{\odot}$ to match the observed flux levels. This luminosity would translate to a stellar radius of {3.13}~R$_{\odot}$, leading to an age of 2-3~Ma, and a stellar mass of about 2~M$_{\odot}$, in disagreement with the observed dynamical mass estimate of 1.3-1.5~M$_{\odot}$. Critically, this high-inclination inner disk model also fails to reproduce the visibilities observed with GRAVITY.}
{The inner dust disk, as traced by the GRAVITY data, is located at a radius in agreement with the dust sublimation radius. An ambiguity remains regarding the respective orientations of the inner and outer disk, coplanar and mildly misaligned, respectively. As our datasets are not contemporary and the star is strongly variable, a deeper investigation will require a dedicated multi-technique observing campaign.}


   \maketitle
% -----------------------
\section{Introduction}
\label{sec:intro}

Because they are the sites of planet formation, the study of protoplanetary disks is one of the primary science drivers for several recent major observing facilities. ALMA, operating at (sub-)millimeter wavelengths \citep{HLTauALMA2015}, and second-generation adaptive-optics imagers operating in the optical/near-infrared(NIR), reach comparably high angular resolutions, providing clear views of the disks at the 10~au scale for the typical star-forming regions (i.e., at $\sim$~140~pc). One of the many outstanding findings to come from such observations is that most of the disks are non-symmetric in nature when angular resolution is improved: resolved disks show gaps, rings, spirals, vortices, and shadows \citep{Andrews2018,Long2018,Benisty2015,deBoer2016,Pohl2017,Benisty2017,Avenhaus2018}. The origin (and universality) of these features is actively debated, given the clear possibility that they are tracing the {dynamical} interaction between a forming planet and its parental disk. However, other mechanisms like snow lines \citep{Zhang2015}, non-ideal MHD effects \citep{Bethune2016}, zonal flows \citep{Loren-Aguilar2015}, and self-induced dust-traps \citep{Gonzalez2015} also have the capacity to produce rings, gaps, and spirals without the need for planets. 

While scattered light imaging in the optical and the NIR with, for example, SPHERE \citep{Beuzit2019} or imaging at (sub-) millimeter wavelengths with ALMA provides detailed views of the outer regions of the disks (i.e., from $\sim$10 to $\sim$500~au), these instruments are not directly sensitive to the very inner regions. Typically, the inner parts of the disks (< 5-10~au) escape from direct view because of the use of a coronagraph with SPHERE, for example.

{ This is perhaps an unfortunate situation, because the techniques most currently used today to search for extrasolar planets, namely those involving radial velocities and transit surveys, are naturally biased towards planets with small periods and thus small separations. Nevertheless, the results from these techniques clearly indicate that planets are frequent around solar-like stars (Clanton and Gaudi, 2016; Baron et al., 2019; Fernandes et al., 2019).}

{ The standard scenario for planet formation is called core accretion \citep{Pollack1996,Rice2003}. The main caveat with the original version of this scenario is that the timescale to form a Jupiter-like planet (at 5 au from the star) is longer than the typical gas-disk lifetime, with the situation becoming even worse for planets located farther away from the central star. Also, there is not enough mass available in-situ to form giant planets closer in, for example in Earth-like orbits or closer in. Nevertheless, several ways of mitigating the timescale and mass problems have been suggested. On one hand, planets can form further out and migrate inward \citep{Alibert2005}. On the other, the streaming instability in the disk midplane \citep{Youdin2005,Johansen2014} and/or pebble accretion \citep{Ormel2010,Lambrechts2012} can significantly speed up the planet formation process. Therefore, knowledge of the disk properties in the midplane is critical, and is even more so in the inner regions.}  
%radial velocity and transit surveys suggest that rocky %planets are likely to form or at least to migrate in %the inner few au's. Indeed, dust is thermally processed %in the innermost disk where magneto-rotational %instability can occur, and dead zone and dust trap can %form. This possible formation of local asymmetries play %a crucial role in the vortex formation and planetesimal %formation \citep{Flock2016,Flock2017}.} 

Complementary to SPHERE and ALMA, long-baseline NIR interferometers provide { the necessary} spatial resolutions of a few milliarcseconds (mas), which is typically a fraction of an astronomical unit at 140~pc. {These instruments allow us to study the dust distribution, composition, and evolution of protoplanetary disks, which provides clues for understanding planet formation. Moreover, a  sharper view of the dust distribution in  the inner astronomical unit is also crucial to better understand  the timescales of disk evolution.} Previous NIR interferometric observations focused mostly on the brightest end of the pre-main sequence stellar distribution for lack of sensitivity. In practice therefore, Herbig Ae/Be stars constitute the bulk of the observed sample \citep[e.g.,][]{Lazareff2017}, with a few  rare exceptions for the brightest T~Tauri stars observed {with the Keck Interferometer \citep{Eisner2007,Eisner2010}, AMBER \citep{Vural2012}, and PIONIER \citep{Anthonioz2015}}. With a significant gain in sensitivity, GRAVITY \citep{GRAVITY2017} provides the opportunity to observe, at high angular resolution, a larger sample of these young, low-mass stars, which appear more compact and fainter in the NIR than the more massive Herbig stars, which is due to their cooler photospheres.

Here, we present the observations of the Solar-like pre-main sequence star RY~Lup obtained with GRAVITY in the K-band. RY~Lup is located at a distance of 158~pc \citep{GAIA-DR2-2018}. Although a range of spectral-type estimates are available in the literature, RY~Lup is now classified as K2 based on X-Shooter spectroscopic data \citep{Alcala2017}. {These latter authors also derived a luminosity of 1.7~L$_{\odot}$} {and a stellar mass of 1.47M$_{\odot}$ (from \citet{Siess2000} tracks)}, and revised the effective temperature to slightly lower values (T$_{\rm eff}$~=~4900~$\pm$~227~K) compared to previous studies { \citep[e.g., 5080~K derived from][]{Frasca2017}. Using the evolutionary tracks of \citet{Baraffe2015}, \citet{Frasca2017} derived a mass of 1.4~M$_{\odot}$ and an age of 10.2~Ma, while most of the young stellar objects of the Lupus association have an age of 1-3 Ma (see their Fig. 6).}

The SED of RY~Lup shows a significant IR excess, as well as a modest UV excess \citep{Evans1982,Gahm1989}. RY~Lup also exhibits strong photometric variability. \citet{Manset2009}, using simultaneous BV polarimetric and UBV photometric observations, showed that the polarization is high (3.0\%) when the star is faint and red (V = 12.0 mag, B-V = 1.3 mag), and low (0.5\%) when it is brighter and bluer (V = 11.0 mag, B-V = 1.1 mag). The photometric and polarimetric variabilities also share a common period of 3.75~d, leading the authors to conclude that these variations are produced by an inclined circumstellar disk that is warped close to the star, where it interacts with the stellar magnetosphere, and that co-rotates with the star.

ALMA observations of RY Lup in the 890-$\mu$m dust continuum \citep{Ansdell2016} provided the first direct indication that the disk of RY~Lup is seen at high inclination. A ring of dust is resolved around RY~Lup, surrounding an inner cavity with a diameter of 0.8" (radius $\sim$ 60~au). The results of the UV-to-NIR study performed by \citep{Arulanantham2018} to probe the details of the cavity and inner disk are fully consistent with the presence of a gas gap within the millimeter(mm)-dust cavity, confirming the pre-transition disk nature of the system. The inclination and the position angle of the disk derived from these ALMA data are 67$^\circ$ and 109$^\circ$, respectively \citep{Francis2020}. The accuracy on these angles was estimated to be 5$^\circ$ by \citet{vanderMarel2018}. The disk dust mass M$_{\rm dust}$ derived from the ALMA 1.3~mm data is 3.7 10$^{-4}$~M$_{\odot}$ and the disk gas mass is 7.1~M$_{\rm Jup}$ \citep{Ansdell2018}{, which is equal to 6.8 10$^{-3}$~M$_{\odot}$}. Using the ALMA $^{13}$CO and C$^{18}$O spectra, \citet{Yen2018} derived a stellar mass of 1.3~$\pm$~0.1~M$_{\odot}$. This is compatible with the estimation of 1.47~$\pm$~0.22~M$_{\odot}$ by \citet{Alcala2017}. \citet{Yen2018} also determined the disk orientation through the velocity-aligned stacking method and obtained an inclination of 55~$\pm$~5$^\circ$ and a position angle of 110$^{+5}_{-10}$deg.

\citet{Langlois2018} presented a scattered light image obtained at H-band with the VLT/SPHERE instrument. The high inclination of the disk is confirmed, and interestingly the surface brightness of the disk reveals no inner gap, as opposed to the ALMA data. This is also a frequent signature of transition disks where the central part of the disk shows a decrease in density but is not completely void of gas and dust, {especially small dust best seen in scattered light. This explains why the transition disk nature of the disk was not recognized in early SED studies \citep[e.g.,][]{Manset2009}.}  

We used GRAVITY guaranteed time observations of RY~Lup {to further constrain the inner geometry} of the  circumstellar environment of RY Lup. We present our observations  in Sect.~\ref{sec:observation} and our modeling approach in Sect.~\ref{sec:modeling}. The results of our modeling are detailed in Sect.~\ref{sec:results} and discussed in Sect.~\ref{sec:interpretation}.

%-------------------------
\section{Observations}
\label{sec:observation}

%This paper uses new data from GRAVITY as well as ancillary interferometric and photometric archival data.

\begin{table*}[t]%
\caption{Near-infrared interferometric observations of RY~Lup.}
\centering
\begin{tabular}{c c c c c c}
\hline
Instrument      & Date  &       {Time} & Telescope configuration        & Spectral band & Calibrator \\ 
\hline
 GRAVITY        &2017-06-11     & {UT02:09-UT04:09}     & UT1-UT2-UT3-UT4 & K-band        & HD 110878 \\ 
\hline
PIONIER & 2011-08-07    & {UT23:48-UT00:23}     & A1-G1-I1-K0           & H-band & HIP~77964, HIP77295 \\
        &2013-05-12     & {UT03:27-UT03:56}     & A1-B2-C1-D0   &        H-band & HIP~78456, HIP~78238 \\
\hline
\end{tabular}
\label{tab:interferometry} 
\end{table*}

\begin{figure*}[t]%
\centering
\includegraphics[width=19.5cm]{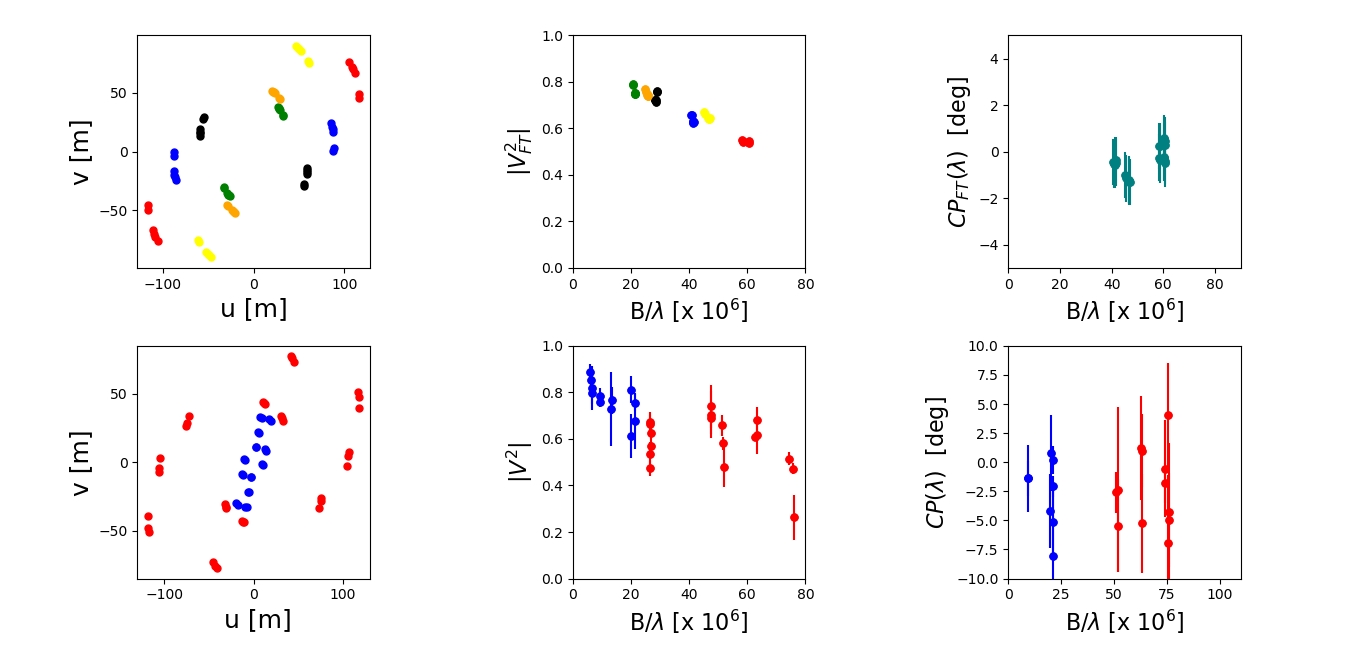}
\caption{K-band GRAVITY (top) and H-band PIONIER (bottom) observations of RY Lup. { Left.} \textit{{(u,v)}} plane. { Middle.} Squared visibilities as a function of the baseline. { Right.} Closure phases as a function of the longest baseline of the telescope triplet. {For GRAVITY observations, each colour corresponds to a baseline, while for PIONIER observations, the different colours illustrate different observing dates and configurations (Table~\ref{tab:interferometry})}}
\label{fig:dataGRAVITY}
\end{figure*}

\subsection{GRAVITY}

%\paragraph{Observations and data reduction}
Within the context of the Young Stellar Objects Large Program of GRAVITY \citep{GRAVITY2017,Eisenhauer2011}, we observed RY~Lup with the Unit Telescopes (UTs) of the VLTI on 2017 June 11 for 2 hours. With the science channels (SCs), we recorded seven data sets at high spectral resolution ($R$~$\sim$~4000) corresponding to ten exposures,  each of 30 seconds integration time. With the fringe tracker (FT), we recorded low-spectral-resolution data (5 channels across the K-band) at a frame rate of about 1~kHz \citep{Lacour2019}. We interlaced our observations with sky exposures and observations of an interferometric calibrator, HD~110878, to retrieve the instrumental transfer function. We used the GRAVITY standard pipeline \citep{Lapeyrere2014} to reduce and calibrate the observations.

%\paragraph{Data}

In this paper, we focus on the low spectral resolution K-band data provided by the FT for probing the inner dust rim in the K-band continuum. The high cadence of the FT frames allows the atmosphere to be frozen during each exposure of 0.85~ms, reducing the smearing effect on the interferometric fringes that can drastically degrade the visibilities. The FT calibrated data and the \textit{{(u,v)}}-plane coverage of our observations are displayed in the top penal of  Fig.\ref{fig:dataGRAVITY}. RY~Lup is partially resolved with the UT baselines, with squared visibilities ranging from 0.55 to 0.8, and we detect no clear departure from centro-symmetry since the closure phases are all in agreement with zero.

\subsection{Complementary data}

\paragraph{VLTI/PIONIER}

RY~Lup was observed in the H-band with the PIONIER instrument of the VLTI \citep{Lebouquin2011}. We used two datasets obtained with two different configurations of the auxiliary telescopes (ATs; Table~\ref{tab:interferometry}). The data were reduced with the standard PIONIER pipeline. RY~Lup is partially resolved in the H-band with visibilities squared between 0.4 and 0.9 (Fig.\ref{fig:dataGRAVITY}-bottom). The visibilities quickly decline at the baselines between zero to a few meters, which is generally interpreted as the contribution of a component that is much more extended than the angular resolution of the interferometer. %As a consequence, its visibility is null at all the spatial frequencies but the null one. Because it partially contributes to the total flux, its net effect is that for very short baselines, the total visibility is smaller than 1. 
{Such a contribution can be interpreted as scattered light, as described in detail in \citet{Pinte2008b} and {as applied in} \citet{Anthonioz2015} who derive a ratio of 0.41 between the scattered light flux and the stellar flux for RY~Lup in their composite model (see their Table 5 and Fig. 2).}

\paragraph{Photometry}

%\fme{J'ai beaucoup change ce paragraphe}\kp{OK pour moi}
We use nonsimultaneous photometric data from the literature  to build the SED (see Table~\ref{tab:photometry}) as well as the Spitzer/IRS spectrum \citep{Chen2016} to cover the mid-IR (MIR) and the silicate features at 10 and 20 microns. In order to minimise the impact of the known stellar variability, we use the X-Shooter spectrum published by \citet{Alcala2017} to provide data over a wide range, from the B- to the K-bands. 

\begin{table}[h]
\caption{Photometric data used to build the SED of RY~Lup.}
\begin{center}
%\begin{tabular}{@{}llll@{}}
\begin{tabular}{llll}
\hline
$\lambda$   & F & $\sigma_{F}$  &  Reference \\ 
 ($\mu$m) & (W.m$^-2$) & (W.m$^-2$) & ~  \\
\hline
%  0.55 & 7.02 $10^{-13}$  &  3.24 $10^{-14}$   & Johnson V   \\
%  1.24 & 1.46 $10^{-12}$  &  3.14 $10^{-14}$   & 2MASS J    \\
%  1.25 & 1.47 $10^{-12}$  &  3.12 $10^{-14}$           & Johnson J    \\ 
%  1.63 & 1.60 $10^{-12}$  & 6.80 $10^{-14}$            & Johnson H    \\
%  1.65 & 1.60 $10^{-12}$  & 7.45 $10^{-14}$            & 2MASS H    \\
%  2.16 & 1.51 $10^{-12}$  & 2.77 $10^{-14}$            & 2MASS K    \\
%  2.19 & 1.45 $10^{-12}$  & 2.73 $10^{-14}$            & Johnson K    \\
 3.40   & 1.33 $10^{-12}$  & 2.6$10^{-13}$                                                      & Johnson L $^{(a)}$   \\
 5.03 & 7.39 $10^{-13}$  & 1.5$10^{-13}$                & Johnson M $^{(a)}$   \\
 11.5& 3.87 $10^{-13}$  & 2.3$10^{-14}$                                                         & IRAS 12 $^{(b)}$   \\
 23.8& 3.55 $10^{-13}$  & 2.5$10^{-14}$                                                                 & IRAS 25 $^{(b)}$   \\
 61.8& 2.72 $10^{-13}$  & 2.4$10^{-14}$                                                                 & IRAS 60 $^{(b)}$   \\
 102& 1.62 $10^{-13}$  & 1.9$10^{-14}$                                                          & IRAS 100 $^{(b)}$   \\
 890 &9.26 $10^{-16}$    & 3.3$10^{-18}$                                                                        & ALMA $^{(c)}$     \\ 
 1330 & 1.94 $10^{-16}$ & 1.5 $10^{-18}$ & ALMA $^{(c)}$ \\
 \hline
\end{tabular}
\label{tab:photometry}
\end{center}
\tablefoot{\noindent $^{(a)}$ \citet{morel1978}; $^{(b)}$ \citet{Abrahamyan2015}; $^{(c)}$ \citet{Ansdell2016};  \citet{Ansdell2018}}
\end{table}

\begin{figure*}[t]%
\centering
\includegraphics[width=15cm]{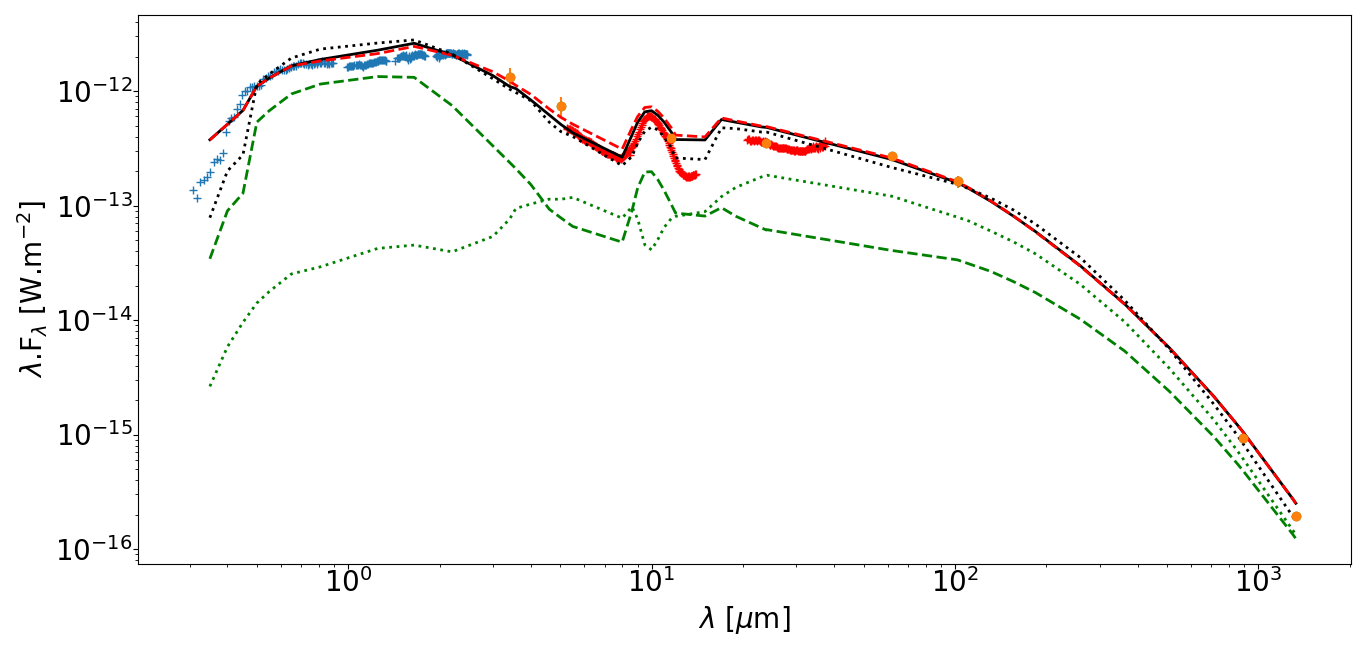}
\caption{{ Spectral energy distributions corresponding to our models of RY Lup: M50 for an inclination of 50$^\circ$ with $r_{in}~=~0.12$~au (black solid line) and $r_{in}~=~0.17$~au (red dashed line), and M70 for an inclination of 70$^\circ$ (black dotted-line)} compared with the photometric measurements (orange circles; Table~\ref{tab:photometry}), the X-Shooter spectrum (blue crosses), and the Spitzer/IRS spectrum (red plus).  { The green lines display the SEDs calculated with the disk parameters of { M50} (dashed line) and of { M70} (dotted line)} models but using the stellar parameters (T$_{\rm eff}$, L$_*$), { and the inclination of 75$^\circ$} used by \citet{Langlois2018}. See text for details.} 
\label{fig:SED}%
\end{figure*}

% http://vizier.u-strasbg.fr/viz-bin/VizieR-4

\section{{A transition disk model for RY~Lup}}
\label{sec:modeling}

We used the radiative transfer code MCFOST to model the SED, the NIR, and sub-mm images, and the NIR H- and K-band interferometric data of RY~Lupi. MCFOST is a 3D continuum radiative transfer code based on the Monte Carlo method \citep[see][for a complete description and benchmarking]{pinte2006,Pinte2009}. 
%The temperature structure and radiation field estimated by the %Monte Carlo runs are used to produce SEDs, images, and %polarisation maps with a ray-tracing method, where the emerging %flux is obtained by calculating the formal solution of the %radiative transfer equation along the rays. 
%\fme{Validez le gras qui suit.} 
{MCFOST includes multiple anisotropic scattering whenever needed with a complete treatment of polarization. It also includes passive dust heating assuming radiative equilibrium, and continuum thermal re-emission. Viscous heating was not considered in our calculations. Very briefly, the code first computes the temperature structure assuming that dust is in radiative equilibrium with the local radiation field. This is done by propagating photon packets, originally emitted by the central star, through a combination of scattering (following Mie theory), absorption, and re-emission events until they exit the computation grid.}

We used a parametric disk model for our calculations. The disk geometry includes a surface density $\Sigma\left(r\right)=\Sigma_0 \left(r/r_0\right)^{-p}$ and a scale height $h\left(r\right) = H_0 \left(r/r_0\right)^{\beta}$~prescription, with $r_0$~=~100~au. During calculations, we ensured that the dust temperature at the inner radius { does not exceed} the sublimation temperature, $T_{\rm sub}$, which we set at 1500~K. { To achieve this we set the position of the inner disk, iteratively if needed, to ensure that the maximum dust temperature does not exceed $T_\mathrm{sub}$}. We considered dust grains with sizes distributed according to the power law $dn\left(a\right) \propto a^{-3.5} da$ and we assumed two different grain size populations, that is, small grains (whose diameters $a$ vary between 0.01 and 1.1~$\mu$m) and large grains ($a$~=~10-1000~$\mu$m). {For simplicity the grains are made of "astronomical silicate" and we used the optical properties published by \citet{Draine1984}.} { In our calculations we assumed that the dust is at local thermodynamic equilibrium, and in radiative equilibrium with the stellar radiation field: at a given position, grains of all sizes have the same temperature, $T_\mathrm{dust}$.} For the photosphere, we adopted a NEXTGEN stellar atmosphere model \citep{Hauschildt1999} with an effective temperature of T$_{\rm eff}$~=~4800~K and a surface gravity of $\log$~g~=4.0.

We distributed the two grain populations in two distinct zones in the disk to describe a typical geometry for transition disks. This choice is guided by the ALMA images showing a ring of emission starting at $\sim$50~au and extending out to $\sim$125~au. This mm emission is dominated by large grains (zone \#2 in our model). On the contrary, the SPHERE NIR image is dominated by scattering and small grains, and shows a disk extending all the way to the center (zone \#1 in our model). Furthermore, following the results obtained for HL~Tau by \citet{Pinte2016} and for a sample of edge-on disks (Villenave et al. 2020, submitted), we distributed the small dust in a flared disk with a standard aspect ratio in zone \#1 (h/r $\sim$ 0.2) but we assumed significant vertical settling for the larger dust in zone \#2. The dust in this zone (i.e., the ALMA ring) is therefore more concentrated toward the midplane with h/r $\sim$ 0.03.
%\fme{les 3 Paragraphes qui suivent sont A VALIDER}

\begin{table*}[t]
\caption{Star and disk parameters of our models of RY~Lup{ , M50, and M70}. {See main text for the definitions and the corresponding references for the fixed values.}}
\centering
\begin{tabular}{l c c c}
\hline
Parameters       & { M50}& { M70}   & Range of explored values       \\
\hline
Effective temperature $T_{\rm eff}$ [K] & 4800& { 4800} & Fixed \\
Stellar luminosity $L_{\star}$  [$L_{\odot}$]   & { 2.5} & { 4.65}    & [1.5-9.5]  \\
Interstellar Extinction $A_{\rm v}$  [mag]         & 0.7& { 1.2}                   & [0.3-{ 1.2}] \\ 
\midrule
{ Disk - Zone \#1} & & \\ 
\midrule
Dust mass $M_{\rm dust}$  [$M_{\odot}$]    & $9.0 \times10^{-5}$  & { $7.0 \times10^{-5}$}    & [$3.0 \times10^{-3}$-$9.0 \times10^{-5}$]   \\

Flaring index $\beta$         & 1.11    & { 1.25}             & [1.015-1.25]    \\

a$_{\rm min}$ [$\mu m$ ]           &  0.01 & { 0.8} &{[0.001- 0.8]} \\
a$_{\rm max}$ [$\mu m$ ]          &  1.1  &  {  1.1} &[1.1-1.7] \\
Scale height $H_0$ [au]$^{(1)}$            & 22 & { 8.5}                   &    [11-23]\\
Inner radius $R_{\rm in}$   [au]     & 0.12          & { 0.29}        &  {[0.08- 0.29]}   \\
Outer radius $R_{\rm out}$   [au]     & 125               & { 125}   &  Fixed   \\
\midrule
{ Disk - Zone \#2} & &  \\ 
\midrule
Dust mass $M_{\rm dust}$  [$M_{\odot}$]    & $2.81 \times10^{-4}$   & { $1.5 \times10^{-4}$}    & [$3.0 \times10^{-3}$-$9.0 \times10^{-5}$]   \\
Flare index $\beta$         & 1.125 & { 1.125}                & { [1.015}-1.25]    \\

a$_{\rm min}$ [$\mu m$ ]           &  10  & { 10} &  [1-15] \\
a$_{\rm max}$ [$\mu m$ ]          &  1000  & { 1000} &[100-1200] \\
Scale height $H_0$ [au]$^{(1)}$            & 3   & { 7}                 &    [2-20]\\
Inner radius $R_{\rm in}$   [au]     & 50          & { 50}        &     Fixed  \\
Outer radius $R_{\rm out}$   [au]    & 125            & { 125}       & Fixed  \\ \midrule

Inclination $i$ [$^{\circ}$]         & 50        & { 70}             &  [0-90]  \\
Position angle $PA$   [$^{\circ}$]$^{(2)}$           & 109& { 109}                   & Fixed \\
{ Maximal temperature at the inner rim $T_{\rm max}$ [K]} & { 1700} & { 1456} & { Computed by MCFOST}\\ 
\midrule
{ Reduced $\rm \chi^{2}_{SED}$ }&   { 23}    &   { 24}   & \_ \\
{ Reduced $\rm \chi^{2}_{V^2}$} &   { 1.58}  &   { 15}   & \_ \\
\hline

\end{tabular}
\label{tab:ModelParam}

\noindent { Notes.} $^{(1)}$ at a reference radius of 100~au. $^{(2)}$ From north to east.
%\fme{ WHERE IS THE OUTER DISK, this would also drive Amax to larger value... etc}
\end{table*}

{Disk modeling can be a large multi-dimensional effort where finding the best-fitting, unique solution is CPU-expensive. Ideally, one would like to identify the best and unique solution to best fit all the data available. For RY~Lup, that would be the SED, the SPHERE and ALMA images (NIR and mm), the IRS spectrum of the silicate features, and the NIR interferometric visibilities at H- and K-bands for example. However, as demonstrated by  \citet[see Fig.~9 and Tab.~5]{Pinte2008} for example, this can  only be achieved by running a very significant number of models. Running models of any individual sets of constraints, that is, the SED alone or the images alone, systematically provides solutions of lesser quality, with broader and shallower probability distribution functions \citep[][for the case of IM Lupi]{Pinte2008}. %\fme{DECIDE to KEEP or IGNORE THIS ONE. }
%\kp{I propose to keep it.} 
{ It is also relevant to stress here that additional constraints, coming a posteriori, can identify different or better solutions that were not indicated by the original modeling effort. This is the case for the models presented in \citet{Pinte2008} for example, where a posteriori CO observations \citep{Panic2009} revealed that the gas disk radius is at least twice that found by \cite{Pinte2008} based on continuum data only. Therefore,} we considered here that finding the ultimate model solution to match all the data available simultaneously { was} beyond the scope of the paper, as the paper focuses primarily on the new results provided by the GRAVITY instrument. We have therefore chosen a simpler approach. First we identified a model that correctly reproduces the SED. The adjustment is made by visual inspection and is not the result of a formal minimization. We focus in particular on  correctly matching the optical/NIR part as this is the wavelength range where, in a second step, we calculate images and visibilities to compare with the VLTI interferometric data. The results from the SED fitting are presented in \S\ref{sec:SED} below. To model the visibilities, the SED model identified in the first step { is then} adjusted, and modified as minimally as possible, to match the NIR interferometric data while maintaining a good match to the SED. The results are presented in \S\ref{sec:visibility}.}

\section{Results}
\label{sec:results}

\subsection{Spectral energy distributiion, circumstellar extinction, and revised stellar luminosity}
\label{sec:SED}

An SED was constructed using the photometric data listed in Table~\ref{tab:photometry}. We {visually tested} several models by spanning different ranges of stellar luminosity (L$_*$), disk dust mass, flaring index ($\beta$), scale height ($H_0$), and inclination ($i$) to broadly match the shape of the SED. {The explored ranges are given in the last column of Table 3.} Well-defined parameters were kept fixed at values derived from previous studies. {Interstellar extinction ($A_v$) was applied a posteriori to the modeled SED. This is in addition to the circumstellar extinction naturally caused by the disk and fully taken into account during calculations.} 
{In this first modeling step, to match the SED only, we fixed the inner radius at{ 0.17~au.}} The disk outer radius R$_{\rm out}$ is set fixed to 125~au \citep{Ansdell2018}, and the total dust mass  M$_{\rm dust}$ is set to 3.7 10$^{-4}$~M$_{\odot}$ {(the sum of the dust mass in the two model disk zones)}, which provides a good match to the SED and the mm-flux. We set the position angle from ALMA observations to 109$^\circ$ \citep{Ansdell2018} and consider a distance of 158~pc \citep{GAIA-DR2-2018}.

{The recent scattered light images presented by \citet{Langlois2018} suggest that the disk reduces the apparent stellar luminosity because of circumstellar extinction, by occulting the photosphere. In our calculations, the luminosity was therefore left as a free parameter. We found that in order to match the observed flux density, a luminosity of $\sim${ 2.5}~L$_{\odot}$ is needed. This is a factor of $\sim${ 1.5} higher than the value of 1.7~L$_{\odot}$ reported by \citet{Alcala2017}. %\fme{Faut ajuster le facteur!  This is 50\% higher than...}
Our revised luminosity translates into a stellar radius of R$_{\star}={ 2.29}\, $R$_{\odot}$, when considering an effective temperature of 4800~K. The circumstellar extinction is calculated self-consistently by the radiative transfer code, which includes the
effects of anisotropic and multiple scattering. An additional small amount of {\sl interstellar extinction}, A$_{\rm v}$ = 0.7~mag, is needed to match the shape of the SED in the optical/NIR (with $R_{\rm v}$~=~3.1). A model that reproduces the overall shape of the SED reasonably well is shown in Figure~\ref{fig:SED}. Its parameters are given in Table 3 { in the column entitled M50 (the inclination of this model is 50$^\circ$}). {As we show in the following section, this inclination and model M50 produce the best agreement with the GRAVITY data.}

\begin{figure*}[t]%
\centering
\includegraphics[width=9cm]{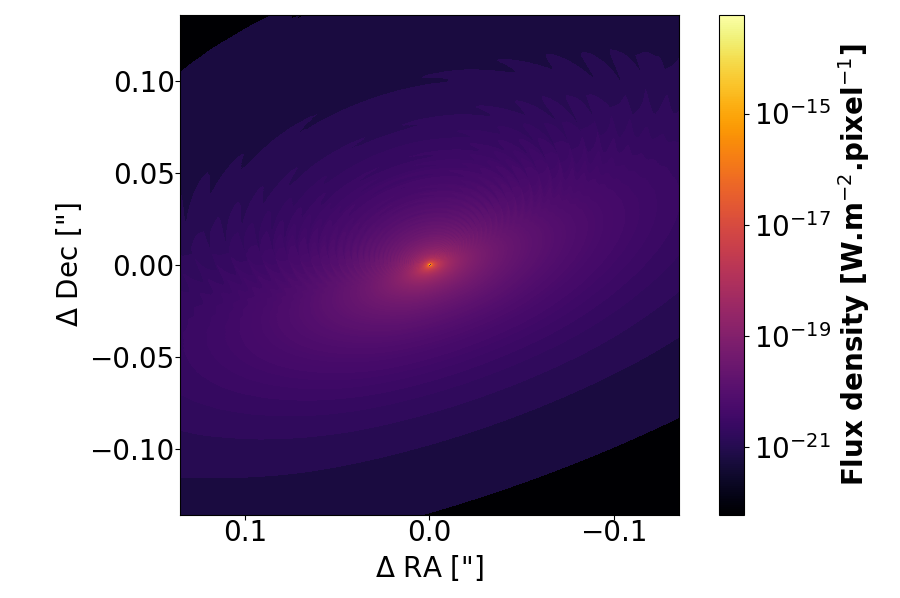}
\includegraphics[width=9cm]{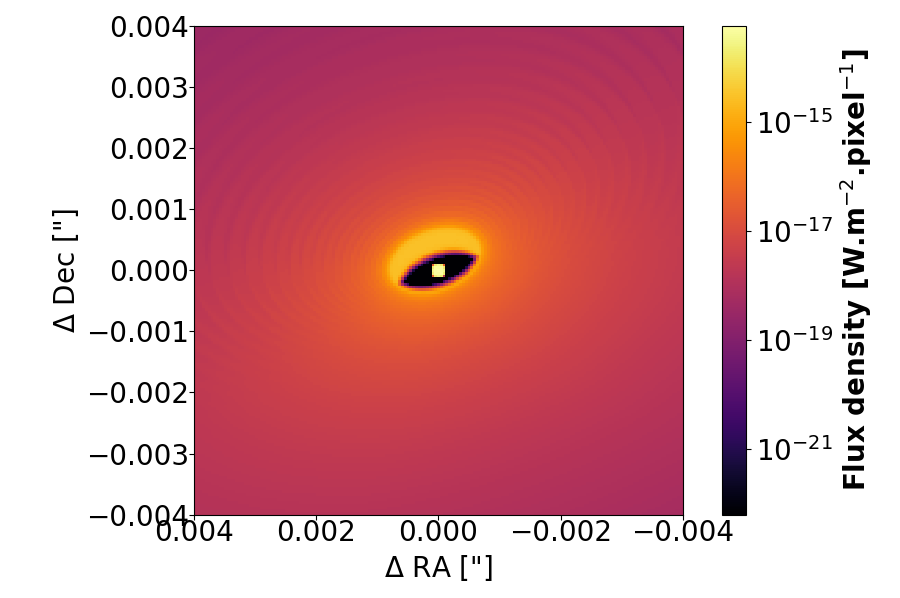}
\includegraphics[width=18.5cm]{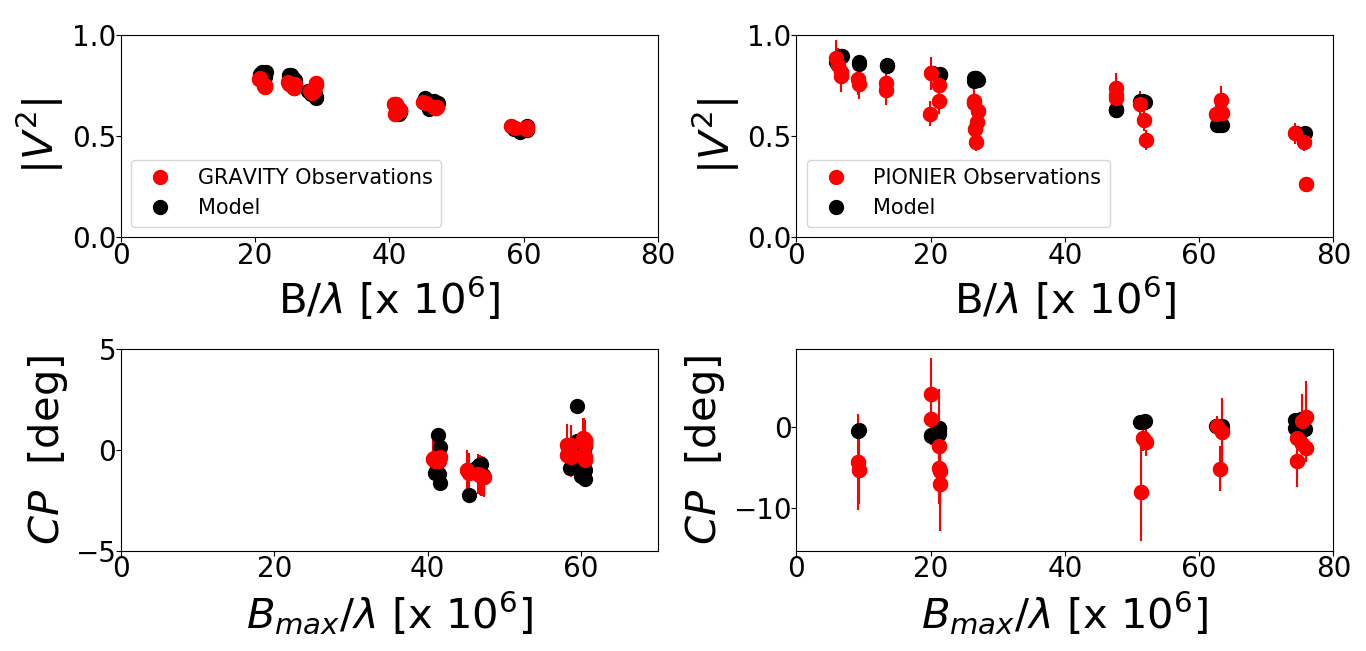}
\caption{{ Top.} Intensity maps produced by MCFOST for our model M50 of RY~Lup in the K band for the whole disk (left) and zoom in at the milliarcsecond scale of the central region (right). { Middle.} Observed squared visibilities (red circles) with GRAVITY (left) and PIONIER (right) compared with the visibilities computed with our model { M50} (black circles). { Bottom.} Observed closure phases (red circles) with GRAVITY (left) and PIONIER (right) compared with the closure phases computed with our model { M50} (black circles).}
\label{fig:M50}
\end{figure*}

{For comparison, we overplotted the SED computed for our disk model { M50}, which has the same geometry and interstellar extinction, but takes the { inclination,} luminosity, and  effective temperature used by \citet{Langlois2018} following \citet{Manset2009}, that is  50$^\circ$,  L$_*~=~2.4~$L$_{\odot}$ (calculated from a stellar radius of 1.72~R$_{\odot}$), and an effective temperature of 5500~K. Clearly, the resulting SED (displayed by the green dashed line in fig.~\ref{fig:SED}) is globally too faint to match the photometric data. %\fme{clarifier ici que meme si c'est M50, la SEd est calculee pour i=70...
%C'est complique de recalculer pour i=75, comme dans langlois? sinon la luminosite etant la meme, il y a la meme quantite d'energie dans le system et ca devrait marcher! SUGGESTION:}
%\kp{Ce que j'ai compris des remarques de François, c'est qu'on recalcule la SED avec nos paramètres de géométrie du disque, les paramètres stellaires de Maud et une inclination de 70 deg. Est-ce correct ? Si c'est bien le cas, effectivement, ça vaudrait le coup de refaire le plot pour i=75. Si c'est le cas, c'est en contradiction avec la légende de la Figure 2.}
{ This is because the model by \citet{Langlois2018} does not include an inner disk. The disk used by these latter authors is a single flat dust ring with a large inner cavity extending out to r~=~18~au, as suggested by ALMA \citep{Ansdell2016}. In that case the photosphere is not occulted by the disk, even though the authors are considering a high inclination of 75\degr, contrary to a case where the inner disk would be located much closer to the star, as in our M50 model.}}

{ However, since the SPHERE and ALMA images of the outer disk suggest an inclination of about 70~$^\circ$ \citep{Langlois2018,Francis2020}, we also investigated the parameter space needed to reproduce the SED at such a high inclination. The SED of this model, that we call M70, is overplotted in black dotted line in Fig.~\ref{fig:SED}. Its parameters are given in the column M70 of Table 3.} { The overall agreement with the SED is similar to, though  slightly poorer than, the one provided by the M50 model { (see the values of reduced $\chi^2$ for the SED fits in Table~\ref{tab:ModelParam})}. Of importance, the flux density in the GRAVITY range is correctly reproduced. Due to the higher inclination of the system, the disk blocks more flux from the star compared to the M50 model. Accordingly, the stellar luminosity of the M70 model has to be higher to match the SED data. We find a luminosity of {4.65}~$L_\odot$, which translates into a stellar radius of {R$_{\star}=3.13\, $R$_{\odot}$} for an effective temperature of 4800~K.}

\subsection{Visibility fitting}
\label{sec:visibility}

%\fme{A Valider:} 
{After identifying a disk model that matches the SED reasonably (Fig.~2), we added one layer of complexity by calculating images and visibilities to verify the capacity of that model to also fit the VLTI interferometric data. We first computed intensity maps in the H- and K-bands with a pixel sampling of 0.05~mas, as illustrated in the top panel of Figure~\ref{fig:M50} for the K-band. This sampling is about 70 times denser than our K-band spatial resolution and 55 times denser than our H-band spatial resolution. These images were then processed by Fourier transform to produce interferometric visibilities. We used the ASPRO2 tool{\footnote{Available at http://www.jmmc.fr/aspro}} provided by JMMC to compute the visibilities with the exact same \textit{{(u,v)}}-coverage as the observations and to take into account the field of view of the telescopes through a convolution with a Gaussian whose half width at half maximum (HWHM) is equal to the point spread function of the telescopes (i.e., 60~mas for the UTs and 250~mas for the ATs).
%R$_{in}$ was moved from 0.1~au to 0.12~au. This demonstrates the interest of NIR interferometry to probe the inner regions of disks. 

\begin{figure*}
\centering
\includegraphics[width=14cm]{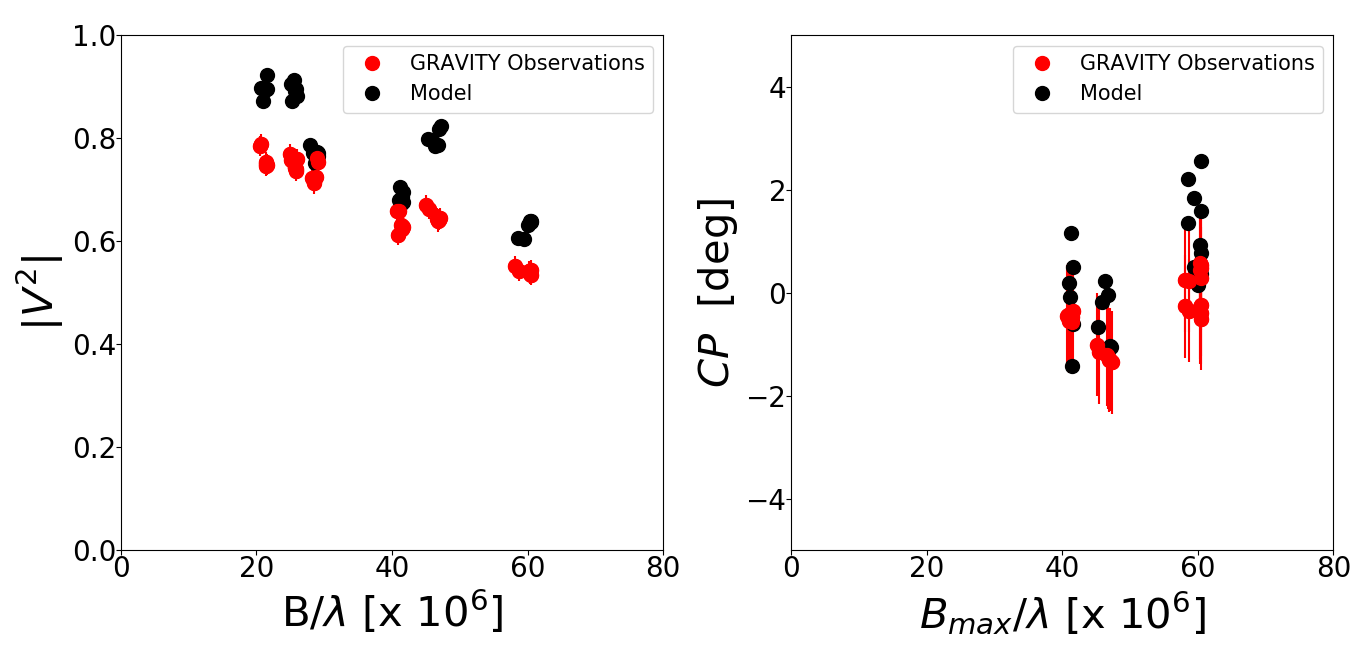}
\caption{ Squared visibilities (left) and closure phases (right) as observed with GRAVITY (red circles) and as computed with our model M70 (black circles).}
\label{fig:M70}%
\end{figure*}

{ For model M50, we needed to slightly adjust the inner disk radius to fine tune the adjustment of the visibility curves, while maintaining a proper match to the SED.} An inclined disk (with an inclination of $\sim$50$^\circ$) and an inner dust rim R$_{in}$ at 0.12~au correctly fits the SED and the NIR interferometric observations (Fig.~\ref{fig:M50}).} { We note that reducing the inner radius to 0.12~au entails an increase in the maximum dust temperature at the inner disk up to 1700~K, slightly beyond to adopted sublimation temperature of 1500~K, but still reasonable.} {The fit of the high-quality GRAVITY {visibilities} is excellent. Our model is also in agreement with the PIONIER data, considering the larger error bars of these datasets, even if the difference between the data and the model is larger. Such a difference could come from the variability of the source because the geometry of the inner disk is expected to be very similar in both the H- and K-bands. {Contemporary interferometric datasets are needed to investigate this in more detail.} We also verified that our { M50} model leads to closure phase signals that are consistent with $\sim$~0$^\circ$, as observed by both GRAVITY and PIONIER { (Fig.~\ref{fig:M50}). This is the case because even if the central part of the disk model is not centro-symmetric, its compactness (1~mas or less) is such that it is only partially resolved by GRAVITY as shown by the squared visibilities higher than 0.55, even at the longest baselines}. The parameters of our final model { M50} are given in Table~\ref{tab:ModelParam}.

%\begin{figure*}[t]%
%\includegraphics[width=7.5cm]{DiskImageFontSize.png}
% \includegraphics[width=11.5cm]{GRAVITY_PIONIERV2FontSize.png}%
%\includegraphics[width=11.5cm]{GRAVITY_PIONIERHKcorr1.png}
%\caption{Zoom in the central region of the intensity map produced by MCFOST for our model of RY~Lup in the K band (left). Observed squared visibilities (red circles) with GRAVITY (middle) and PIONIER (right) compared with the visibilities computed with our model (black circles). }%
%\label{fig:vis2resu}%
%\end{figure*}

%\begin{figure*}[t]%
%\centering
% \includegraphics[width=12.5cm]{SEDFontSize.png}
%\includegraphics[width=15.5cm]{ClosurPhaseModelData1.png}
%
%\caption{\textbf{Measured closure phases (red circles) of RY Lup with GRAVITY (left) and PIONIER (right) compared with the closure phases from the model (black circles).}}
%\label{fig:cp}%
%\end{figure*}

{ For the M70 model, we failed to correctly  reproduce the GRAVITY visibilities (Fig~\ref{fig:M70}), even when changing the inner rim position which is about three times further away from the star than for the M50 model due to the higher luminosity. In particular, the modeled visibilities along three baselines exhibiting a position angle of 150-160$^\circ$ (in green, orange, and yellow in Fig. 1) remain systematically larger than those measured during observations. { See the values of reduced $\chi^2$ for the visibility fits in Table~\ref{tab:ModelParam}).}}

\section{Discussion and concluding remarks}
\label{sec:interpretation}

%\fme{J'ai tout change, A VALIDER}
{Taking into account the dimming of the photospheric flux caused by the extinction from the inclined disk, we derive a revised luminosity of { {2.5} L$_{\odot}$ for an inclination of 50$^\circ$ (M50)}. For comparison, \citet{Alcala2017} report a luminosity of 1.7~L$_{\odot}$ for a similar effective temperature.} {With the effective temperature and luminosity we estimated { for M50}, RY~Lup registers in the Hertzsprung-Russell diagram immediately outside the range calculated by \citet{Baraffe2015} (see Fig. 6 of \citet{Frasca2017}, in-between the 1~ and 3~Ma isochrones). These latter calculations stop at 1.4~M$_{\odot}$, implying a small extrapolation, but such an extrapolation would indicate a mass slightly above 1.4~M$_{\odot}$ and an age of $\sim$~{ 2~Ma} for RY Lup { for M50}. Using the \citet{Siess2000} tracks for Lupus, this would { correspond to a $\sim$~1.6~M$_{\odot}$ and 3-4~Ma} pre-main sequence star (see, e.g., Fig.~2 of \citet{Alcala2017}). A comparison with older pre-main sequence tracks by \citet{DAntona1994} converts into a $\sim$~{ 1.3}~M$_{\odot}$ and $\sim$~1~Ma pre-main sequence object, using their CM convection models %\fme{Please check that one, we changed our luminosity, so these numbers will changed. I checked the other 2}.
%\kp{Done}
In all cases, this brings the age of RY Lup (about 2.0~Ma) more in line with the age of the Lupus association, in opposition to the estimations of  for example 10.2~Ma by \citet{Frasca2017} and 12~Ma by \citet{Manset2009}. This discrepancy can be explained by the systematic underestimation of the luminosity caused by not taking into account the circumstellar extinction due to the disk, and only assuming a mild interstellar extinction of the photosphere in agreement with the observed photospheric colors. The underluminous nature of highly inclined disk systems in Lupus is discussed in \citet{Frasca2017}, although not specifically for RY~Lup. Other similar cases of underluminous, highly inclined disk systems in Lupus, whose ages {are drastically higher than the age of the Lupus association}, include MY~Lupi, Sz 112, and Par-Lup 3-4 \citep{Alcala2017}. }

{{ For M50,} the evolutionary models mentioned above also predict masses that are in agreement with observations. The estimates range from 1.47~$\pm$~0.22~$M_{\odot}$ to 1.3~$\pm$~0.1~$M_{\odot}$ \citep[][respectively]{Alcala2017,Yen2018}}, the latter being a "dynamical mass" estimated from the observed rotation of the gas disk. 
%For completeness, we note however that the stellar mass of $\sim$~2~$M_{\odot}$, estimated from \citet{Siess2000} tracks, is above the current estimated values.\fme{remove last sentence, Siess gives 1.6msun with the new luminosity} 
{ For the model with an inclination of 70$^\circ$ (M70), the derived luminosity is higher ({4.65}~L$_{\odot}$) and corresponds to a mass larger than 2~M$_{\odot}$ using the Siess tracks, which is well above the observed estimates.} At this stage, { we need to recall that our favored model for the system (M50)} does not include a detailed modeling of the scattered light image, and in particular the solution for the disk scale height is driven by the SED, and it is therefore possible (if not probable) that the stellar luminosity we derived, although improved with respect to previous estimates, still requires refinement based on a much more thorough analysis of all the data available. Given the intrinsic  large variability of the star, such refinement is well beyond the scope of this paper.}

%\fme{A VALIDER The fundamental stellar parameters we inferred are consistent with an earlier age for %RY Lupi than previously estimated. of about 0.5-1.0~Ma, when considering a stellar radius of %3.3~$R_\odot$, a stellar mass \fme{of $\sim$1.3~$M_\odot$}, and an effective temperature of 4800~K, %and using the evolutionary tracks of \citet{Baraffe2015} and \citet{Siess2000}. Such an age is more %in agreement with the 1-2 Ma estimated age of the Lupus association than the previous determinations %(for instance 10.2~Ma in \citet{Frasca2017} or 12~Ma in \citet{Manset2009}). 

{ Within the range of parameters explored in Table 3, we could not find another disk model that was able to reproduce both the SED and the visibilities properly.} Indeed, many parameters are well constrained either by the SED (e.g., the stellar luminosity and the extinction, the disk dust mass, and flaring) and/or by the interferometric data (the inner radius). 
%\fme{J'ai fait peter une phrase ici, voir les commentaires dans le .tex}
%As an example, the position of the inner rim is thus very well %constrained at 0.12 $\pm$ 0.01~au \kp{Valeurs à confirmer} by %the interferometric data.

{It is worth comparing the inclination of the dusty disk we derive to match the SED and interferometric data with the large-scale disk images provided by SPHERE and ALMA. Our models with inclinations larger than 50$^\circ$ produce too much extinction in the SED. %(see the dotted green line in Fig.~2). 
More extinction would force the stellar luminosity to higher values than { 2.5~L$_{\odot}$}, making the object more massive and in tension with current mass and age estimates{, as demonstrated with the M70 model}. A disk inclination of 50~$^\circ$ is in agreement with the inclination determined from the ALMA CO spectra analysed by \citet{Yen2018} (55~$\pm$~5$^\circ$). However, the inclinations derived from the aspect ratio of the large-scale outer disk imaged by SPHERE \citep[70~$^\circ$;][]{Langlois2018} and ALMA \citep[67~$\pm$~5$^\circ$;][]{Francis2020} are marginally discrepant with our model. Finally, { the comparison of the GRAVITY interferometric data with model M70 clearly shows that this model is { unable to reproduce the visibility curve in detail}, in particular for the three baselines that directly probe the front of the inner rim due to their orientations. 

{ In all of the models presented above the inner disk is sharp, with a clear density truncation in the radial direction at the inner radius. We explored a more realistic geometry, although parametric, for the inner disk by making the inner wall rounder (larger $\mathrm R_{in}$ with increasing |z| distance). The jump in density was also made smoother by using a steep exponential taper in the radial direction at the inner radius.
%
%Even when trying to smooth this front to mimic an inner %rim \kp{Explain what has been changed in the disk %model}\fme{c'est discute nulle part aillaurs, les %parametres ne sont pas donnes dans la table 3. Je %laisserait tomber ces phrases.} \kp{compte tenu de ce %qu'on dit avant, on pourrait penser qu'en lissant le %front, on devrait pouvoir ajuster mieux les visibilités %le long des 3 bases alignées. Il me semble donc utile %de répondre dès à présent à cette objection que %pourrait faire le referee}, 
%
Even with this parametrization of the density at the inner rim we failed to correctly fit the visibility curve at all spatial frequencies.} Our model considers only a single inclination for the inner and outer parts of the circumstellar disk for now, and  we have not yet investigated a potential warp or misalignment to reconcile the observed inclinations of the inner and outer disks.} {To go further, it would be necessary to improve the details of the vertical structure of the disk. A vertically thinner disk would allow for example to accommodate a larger inclination while maintaining the circumstellar extinction at the current level.}

{\citet{Langlois2018} reports the presence of spiral arms in the H-band scattered light image. If a planet is driving these spirals, then the radial and vertical disk structure can be dynamically altered by the forming planetary companion, leading among other effects to a misalignment between the inner and outer disks that could also reconcile the inclination measurements of the inner and outer disks. Such a misalignment is also invoked to explain the shadows observed in the SPHERE images \citep{Marino2015}. Complementary high-angular-resolution techniques covering different spectral ranges (e.g., GRAVITY, SPHERE, ALMA) allow such inclination differences in protoplanetary disks to be  to investigated provided the interferometric \textit{{(u,v)}}-coverage is high enough to produce a high-fidelity image of the inner disk, which is not the case for our observations of RY~Lup. Within this context, the sensitivity gain brought by the NAOMI adaptive optics systems on the ATs \citep{Woillez2019} is important for expanding the \textit{{(u,v)}}-coverage and also for improving the measurement accuracy to look for variability and/or to better constrain the scattering phenomena that can be investigated through interferometric observations at very short baselines. Additional and complementary data, also including the interferometric instrument MATISSE \citep{Lopez2018} operating in the MIR range, and probing the intermediate scales between the very innermost regions (<~1~au) probed by NIR interferometry and the few tens of astronomical units probed by ALMA, would allow us to constrain the more complex features of this dusty disk. For this purpose, due to the variability of the source, a multi-technique campaign of simultaneous observations would be required.}

\begin{acknowledgements}
%\fme{Should we thank anyone for the XSHOOTER spectrum?}
GRAVITY was developed via a collaboration of the Max Planck Institute for Extraterrestrial Physics, LESIA of Paris Observatory and IPAG of Universit\'e Grenoble Alpes/CNRS, the Max Planck Institute for Astronomy, the University of Cologne, the Centro Multidisciplinar de Astrofisica Lisbon and Porto, and the European Southern Observatory. The authors warmly thank C. Manara for providing the X-shooter spectrum. Y.-I.B. acknowledges support by Science Foundation Ireland under under Grant No. 18/SIRG/5597. K.P. acknowledges funding from LABEX OSUG@2020 (Investissements d'avenir ANR10LABX56). F.M. acknowledges funding from ANR of France under contract number ANR-16-CE31-0013. TH acknowledges support from the European Research Council under the Horizon 2020 Framework Program via the ERC Advanced Grant "Origins" (grant number 832428).A.A., M.F. and P.G. were supported by Funda\c{c}\~{a}o para a Ci\^{e}ncia e a Tecnologia, with grants reference UIDB/00099/2020 and SFRH/BSAB/142940/2018. This work was supported by the "Programme National de Physique Stellaire" (PNPS) of CNRS/INSU co-funded by CEA and CNES. This research has made use of the NASA Astrophysics Data System and of the Jean-Marie Mariotti Center \texttt{Aspro} service\footnote{Available at http://www.jmmc.fr/aspro}.\\

\end{acknowledgements}
\bibliographystyle{aa} % style aa.bst
\bibliography{rylupi.bib}

\begin{thebibliography}{60}
\expandafter\ifx\csname natexlab\endcsname\relax\def\natexlab#1{#1}\fi

\bibitem[{{Abrahamyan} {et~al.}(2015){Abrahamyan}, {Mickaelian}, \&
  {Knyazyan}}]{Abrahamyan2015}
{Abrahamyan}, H.~V., {Mickaelian}, A.~M., \& {Knyazyan}, A.~V. 2015, Astronomy
  and Computing, 10, 99

\bibitem[{{Alcal{\'a}} {et~al.}(2017){Alcal{\'a}}, {Manara}, {Natta}, {Frasca},
  {Testi}, {Nisini}, {Stelzer}, {Williams}, {Antoniucci}, {Biazzo}, {Covino},
  {Esposito}, {Getman}, \& {Rigliaco}}]{Alcala2017}
{Alcal{\'a}}, J.~M., {Manara}, C.~F., {Natta}, A., {et~al.} 2017, \aap, 600,
  A20

\bibitem[{{Alibert} {et~al.}(2005){Alibert}, {Mordasini}, {Benz}, \&
  {Winisdoerffer}}]{Alibert2005}
{Alibert}, Y., {Mordasini}, C., {Benz}, W., \& {Winisdoerffer}, C. 2005, \aap,
  434, 343

\bibitem[{{ALMA Partnership} {et~al.}(2015){ALMA Partnership}, {Brogan},
  {P{\'e}rez}, {Hunter}, {Dent}, {Hales}, {Hills}, {Corder}, {Fomalont},
  {Vlahakis}, {Asaki}, {Barkats}, {Hirota}, {Hodge}, {Impellizzeri}, {Kneissl},
  {Liuzzo}, {Lucas}, {Marcelino}, {Matsushita}, {Nakanishi}, {Phillips},
  {Richards}, {Toledo}, {Aladro}, {Broguiere}, {Cortes}, {Cortes}, {Espada},
  {Galarza}, {Garcia-Appadoo}, {Guzman-Ramirez}, {Humphreys}, {Jung}, {Kameno},
  {Laing}, {Leon}, {Marconi}, {Mignano}, {Nikolic}, {Nyman}, {Radiszcz},
  {Remijan}, {Rod{\'o}n}, {Sawada}, {Takahashi}, {Tilanus}, {Vila Vilaro},
  {Watson}, {Wiklind}, {Akiyama}, {Chapillon}, {de Gregorio-Monsalvo}, {Di
  Francesco}, {Gueth}, {Kawamura}, {Lee}, {Nguyen Luong}, {Mangum}, {Pietu},
  {Sanhueza}, {Saigo}, {Takakuwa}, {Ubach}, {van Kempen}, {Wootten},
  {Castro-Carrizo}, {Francke}, {Gallardo}, {Garcia}, {Gonzalez}, {Hill},
  {Kaminski}, {Kurono}, {Liu}, {Lopez}, {Morales}, {Plarre}, {Schieven},
  {Testi}, {Videla}, {Villard}, {Andreani}, {Hibbard}, \&
  {Tatematsu}}]{HLTauALMA2015}
{ALMA Partnership}, {Brogan}, C.~L., {P{\'e}rez}, L.~M., {et~al.} 2015, \apjl,
  808, L3

\bibitem[{{Andrews} {et~al.}(2018){Andrews}, {Huang}, {P{\'e}rez}, {Isella},
  {Dullemond}, {Kurtovic}, {Guzm{\'a}n}, {Carpenter}, {Wilner}, {Zhang}, {Zhu},
  {Birnstiel}, {Bai}, {Benisty}, {Hughes}, {{\"O}berg}, \&
  {Ricci}}]{Andrews2018}
{Andrews}, S.~M., {Huang}, J., {P{\'e}rez}, L.~M., {et~al.} 2018, \apjl, 869,
  L41

\bibitem[{{Ansdell} {et~al.}(2018){Ansdell}, {Williams}, {Trapman}, {van
  Terwisga}, {Facchini}, {Manara}, {van der Marel}, {Miotello}, {Tazzari},
  {Hogerheijde}, {Guidi}, {Testi}, \& {van Dishoeck}}]{Ansdell2018}
{Ansdell}, M., {Williams}, J.~P., {Trapman}, L., {et~al.} 2018, \apj, 859, 21

\bibitem[{{Ansdell} {et~al.}(2016){Ansdell}, {Williams}, {van der Marel},
  {Carpenter}, {Guidi}, {Hogerheijde}, {Mathews}, {Manara}, {Miotello},
  {Natta}, {Oliveira}, {Tazzari}, {Testi}, {van Dishoeck}, \& {van
  Terwisga}}]{Ansdell2016}
{Ansdell}, M., {Williams}, J.~P., {van der Marel}, N., {et~al.} 2016, \apj,
  828, 46

\bibitem[{{Anthonioz} {et~al.}(2015){Anthonioz}, {M{\'e}nard}, {Pinte}, {Le
  Bouquin}, {Benisty}, {Thi}, {Absil}, {Duch{\^e}ne}, {Augereau}, {Berger},
  {Casassus}, {Duvert}, {Lazareff}, {Malbet}, {Millan-Gabet}, {Schreiber},
  {Traub}, \& {Zins}}]{Anthonioz2015}
{Anthonioz}, F., {M{\'e}nard}, F., {Pinte}, C., {et~al.} 2015, \aap, 574, A41

\bibitem[{{Arulanantham} {et~al.}(2018){Arulanantham}, {France}, {Hoadley},
  {Manara}, {Schneider}, {Alcal{\'a}}, {Banzatti}, {G{\"u}nther}, {Miotello},
  {van der Marel}, {van Dishoeck}, {Walsh}, \& {Williams}}]{Arulanantham2018}
{Arulanantham}, N., {France}, K., {Hoadley}, K., {et~al.} 2018, \apj, 855, 98

\bibitem[{{Avenhaus} {et~al.}(2018){Avenhaus}, {Quanz}, {Garufi}, {Perez},
  {Casassus}, {Pinte}, {Bertrang}, {Caceres}, {Benisty}, \&
  {Dominik}}]{Avenhaus2018}
{Avenhaus}, H., {Quanz}, S.~P., {Garufi}, A., {et~al.} 2018, \apj, 863, 44

\bibitem[{{Baraffe} {et~al.}(2015){Baraffe}, {Homeier}, {Allard}, \&
  {Chabrier}}]{Baraffe2015}
{Baraffe}, I., {Homeier}, D., {Allard}, F., \& {Chabrier}, G. 2015, \aap, 577,
  A42

\bibitem[{{Benisty} {et~al.}(2015){Benisty}, {Juhasz}, {Boccaletti},
  {Avenhaus}, {Milli}, {Thalmann}, {Dominik}, {Pinilla}, {Buenzli}, {Pohl},
  {Beuzit}, {Birnstiel}, {de Boer}, {Bonnefoy}, {Chauvin}, {Christiaens},
  {Garufi}, {Grady}, {Henning}, {Huelamo}, {Isella}, {Langlois}, {M{\'e}nard},
  {Mouillet}, {Olofsson}, {Pantin}, {Pinte}, \& {Pueyo}}]{Benisty2015}
{Benisty}, M., {Juhasz}, A., {Boccaletti}, A., {et~al.} 2015, \aap, 578, L6

\bibitem[{{Benisty} {et~al.}(2017){Benisty}, {Stolker}, {Pohl}, {de Boer},
  {Lesur}, {Dominik}, {Dullemond}, {Langlois}, {Min}, {Wagner}, {Henning},
  {Juhasz}, {Pinilla}, {Facchini}, {Apai}, {van Boekel}, {Garufi}, {Ginski},
  {M{\'e}nard}, {Pinte}, {Quanz}, {Zurlo}, {Boccaletti}, {Bonnefoy}, {Beuzit},
  {Chauvin}, {Cudel}, {Desidera}, {Feldt}, {Fontanive}, {Gratton}, {Kasper},
  {Lagrange}, {LeCoroller}, {Mouillet}, {Mesa}, {Sissa}, {Vigan}, {Antichi},
  {Buey}, {Fusco}, {Gisler}, {Llored}, {Magnard}, {Moeller-Nilsson}, {Pragt},
  {Roelfsema}, {Sauvage}, \& {Wildi}}]{Benisty2017}
{Benisty}, M., {Stolker}, T., {Pohl}, A., {et~al.} 2017, \aap, 597, A42

\bibitem[{{B{\'e}thune} {et~al.}(2016){B{\'e}thune}, {Lesur}, \&
  {Ferreira}}]{Bethune2016}
{B{\'e}thune}, W., {Lesur}, G., \& {Ferreira}, J. 2016, \aap, 589, A87

\bibitem[{{Beuzit} {et~al.}(2019){Beuzit}, {Vigan}, {Mouillet}, {Dohlen},
  {Gratton}, {Boccaletti}, {Sauvage}, {Schmid}, {Langlois}, {Petit},
  {Baruffolo}, {Feldt}, {Milli}, {Wahhaj}, {Abe}, {Anselmi}, {Antichi},
  {Barette}, {Baudrand}, {Baudoz}, {Bazzon}, {Bernardi}, {Blanchard}, {Brast},
  {Bruno}, {Buey}, {Carbillet}, {Carle}, {Cascone}, {Chapron}, {Charton},
  {Chauvin}, {Claudi}, {Costille}, {De Caprio}, {de Boer}, {Delboulb{\'e}},
  {Desidera}, {Dominik}, {Downing}, {Dupuis}, {Fabron}, {Fantinel}, {Farisato},
  {Feautrier}, {Fedrigo}, {Fusco}, {Gigan}, {Ginski}, {Girard}, {Giro},
  {Gisler}, {Gluck}, {Gry}, {Henning}, {Hubin}, {Hugot}, {Incorvaia}, {Jaquet},
  {Kasper}, {Lagadec}, {Lagrange}, {Le Coroller}, {Le Mignant}, {Le Ruyet},
  {Lessio}, {Lizon}, {Llored}, {Lundin}, {Madec}, {Magnard}, {Marteaud},
  {Martinez}, {Maurel}, {M{\'e}nard}, {Mesa}, {M{\"o}ller-Nilsson}, {Moulin},
  {Moutou}, {Orign{\'e}}, {Parisot}, {Pavlov}, {Perret}, {Pragt}, {Puget},
  {Rabou}, {Ramos}, {Reess}, {Rigal}, {Rochat}, {Roelfsema}, {Rousset}, {Roux},
  {Saisse}, {Salasnich}, {Santambrogio}, {Scuderi}, {Segransan}, {Sevin},
  {Siebenmorgen}, {Soenke}, {Stadler}, {Suarez}, {Tiph{\`e}ne}, {Turatto},
  {Udry}, {Vakili}, {Waters}, {Weber}, {Wildi}, {Zins}, \&
  {Zurlo}}]{Beuzit2019}
{Beuzit}, J.~L., {Vigan}, A., {Mouillet}, D., {et~al.} 2019, \aap, 631, A155

\bibitem[{{Chen} {et~al.}(2016){Chen}, {Luo}, {Liu}, \& {Jiang}}]{Chen2016}
{Chen}, R., {Luo}, A., {Liu}, J., \& {Jiang}, B. 2016, \aj, 151, 146

\bibitem[{{D'Antona} \& {Mazzitelli}(1994)}]{DAntona1994}
{D'Antona}, F. \& {Mazzitelli}, I. 1994, \apjs, 90, 467

\bibitem[{{de Boer} {et~al.}(2016){de Boer}, {Salter}, {Benisty}, {Vigan},
  {Boccaletti}, {Pinilla}, {Ginski}, {Juhasz}, {Maire}, {Messina}, {Desidera},
  {Cheetham}, {Girard}, {Wahhaj}, {Langlois}, {Bonnefoy}, {Beuzit}, {Buenzli},
  {Chauvin}, {Dominik}, {Feldt}, {Gratton}, {Hagelberg}, {Isella}, {Janson},
  {Keller}, {Lagrange}, {Lannier}, {Menard}, {Mesa}, {Mouillet}, {Mugrauer},
  {Peretti}, {Perrot}, {Sissa}, {Snik}, {Vogt}, {Zurlo}, \& {SPHERE
  Consortium}}]{deBoer2016}
{de Boer}, J., {Salter}, G., {Benisty}, M., {et~al.} 2016, \aap, 595, A114

\bibitem[{{Draine} \& {Lee}(1984)}]{Draine1984}
{Draine}, B.~T. \& {Lee}, H.~M. 1984, \apj, 285, 89

\bibitem[{{Eisenhauer} {et~al.}(2011){Eisenhauer}, {Perrin}, {Brandner},
  {Straubmeier}, {Perraut}, {Amorim}, {Sch{\"o}ller}, {Gillessen}, {Kervella},
  {Benisty}, {Araujo-Hauck}, {Jocou}, {Lima}, {Jakob}, {Haug}, {Cl{\'e}net},
  {Henning}, {Eckart}, {Berger}, {Garcia}, {Abuter}, {Kellner}, {Paumard},
  {Hippler}, {Fischer}, {Moulin}, {Villate}, {Avila}, {Gr{\"a}ter}, {Lacour},
  {Huber}, {Wiest}, {Nolot}, {Carvas}, {Dorn}, {Pfuhl}, {Gendron}, {Kendrew},
  {Yazici}, {Anton}, {Jung}, {Thiel}, {Choquet}, {Klein}, {Teixeira}, {Gitton},
  {Moch}, {Vincent}, {Kudryavtseva}, {Str{\"o}bele}, {Sturm}, {F{\'e}dou},
  {Lenzen}, {Jolley}, {Kister}, {Lapeyr{\`e}re}, {Naranjo}, {Lucuix},
  {Hofmann}, {Chapron}, {Neumann}, {Mehrgan}, {Hans}, {Rousset}, {Ramos},
  {Suarez}, {Lederer}, {Reess}, {Rohloff}, {Haguenauer}, {Bartko}, {Sevin},
  {Wagner}, {Lizon}, {Rabien}, {Collin}, {Finger}, {Davies}, {Rouan},
  {Wittkowski}, {Dodds-Eden}, {Ziegler}, {Cassaing}, {Bonnet}, {Casali},
  {Genzel}, \& {Lena}}]{Eisenhauer2011}
{Eisenhauer}, F., {Perrin}, G., {Brandner}, W., {et~al.} 2011, The Messenger,
  143, 16

\bibitem[{{Eisner} {et~al.}(2007){Eisner}, {Hillenbrand}, {White}, {Bloom},
  {Akeson}, \& {Blake}}]{Eisner2007}
{Eisner}, J.~A., {Hillenbrand}, L.~A., {White}, R.~J., {et~al.} 2007, \apj,
  669, 1072

\bibitem[{{Eisner} {et~al.}(2010){Eisner}, {Monnier}, {Woillez}, {Akeson},
  {Millan-Gabet}, {Graham}, {Hillenbrand}, {Pott}, {Ragland}, \&
  {Wizinowich}}]{Eisner2010}
{Eisner}, J.~A., {Monnier}, J.~D., {Woillez}, J., {et~al.} 2010, \apj, 718, 774

\bibitem[{{Evans} {et~al.}(1982){Evans}, {Bode}, {Whittet}, {Davies},
  {Kilkenny}, \& {Baines}}]{Evans1982}
{Evans}, A., {Bode}, M.~F., {Whittet}, D.~C.~B., {et~al.} 1982, \mnras, 199,
  37P

\bibitem[{{Francis} \& {van der Marel}(2020)}]{Francis2020}
{Francis}, L. \& {van der Marel}, N. 2020, arXiv e-prints, arXiv:2003.00079

\bibitem[{{Frasca} {et~al.}(2017){Frasca}, {Biazzo}, {Alcal{\'a}}, {Manara},
  {Stelzer}, {Covino}, \& {Antoniucci}}]{Frasca2017}
{Frasca}, A., {Biazzo}, K., {Alcal{\'a}}, J.~M., {et~al.} 2017, \aap, 602, A33

\bibitem[{{Gahm} {et~al.}(1989){Gahm}, {Fischerstrom}, {Liseau}, \&
  {Lindroos}}]{Gahm1989}
{Gahm}, G.~F., {Fischerstrom}, C., {Liseau}, R., \& {Lindroos}, K.~P. 1989,
  \aap, 211, 115

\bibitem[{{Gaia Collaboration} {et~al.}(2018){Gaia Collaboration}, {Brown},
  {Vallenari}, {Prusti}, {de Bruijne}, {Babusiaux}, {Bailer-Jones}, {Biermann},
  {Evans}, {Eyer}, {Jansen}, {Jordi}, {Klioner}, {Lammers}, {Lindegren},
  {Luri}, {Mignard}, {Panem}, {Pourbaix}, {Randich}, {Sartoretti}, {Siddiqui},
  {Soubiran}, {van Leeuwen}, {Walton}, {Arenou}, {Bastian}, {Cropper},
  {Drimmel}, {Katz}, {Lattanzi}, {Bakker}, {Cacciari}, {Casta{\~n}eda},
  {Chaoul}, {Cheek}, {De Angeli}, {Fabricius}, {Guerra}, {Holl}, {Masana},
  {Messineo}, {Mowlavi}, {Nienartowicz}, {Panuzzo}, {Portell}, {Riello},
  {Seabroke}, {Tanga}, {Th{\'e}venin}, {Gracia-Abril}, {Comoretto},
  {Garcia-Reinaldos}, {Teyssier}, {Altmann}, {Andrae}, {Audard},
  {Bellas-Velidis}, {Benson}, {Berthier}, {Blomme}, {Burgess}, {Busso},
  {Carry}, {Cellino}, {Clementini}, {Clotet}, {Creevey}, {Davidson}, {De
  Ridder}, {Delchambre}, {Dell'Oro}, {Ducourant},
  {Fern{\'a}ndez-Hern{\'a}ndez}, {Fouesneau}, {Fr{\'e}mat}, {Galluccio},
  {Garc{\'\i}a-Torres}, {Gonz{\'a}lez-N{\'u}{\~n}ez}, {Gonz{\'a}lez-Vidal},
  {Gosset}, {Guy}, {Halbwachs}, {Hambly}, {Harrison}, {Hern{\'a}ndez},
  {Hestroffer}, {Hodgkin}, {Hutton}, {Jasniewicz}, {Jean-Antoine-Piccolo},
  {Jordan}, {Korn}, {Krone-Martins}, {Lanzafame}, {Lebzelter}, {L{\"o}ffler},
  {Manteiga}, {Marrese}, {Mart{\'\i}n-Fleitas}, {Moitinho}, {Mora}, {Muinonen},
  {Osinde}, {Pancino}, {Pauwels}, {Petit}, {Recio-Blanco}, {Richards},
  {Rimoldini}, {Robin}, {Sarro}, {Siopis}, {Smith}, {Sozzetti}, {S{\"u}veges},
  {Torra}, {van Reeven}, {Abbas}, {Abreu Aramburu}, {Accart}, {Aerts},
  {Altavilla}, {{\'A}lvarez}, {Alvarez}, {Alves}, {Anderson}, {Andrei},
  {Anglada Varela}, {Antiche}, {Antoja}, {Arcay}, {Astraatmadja}, {Bach},
  {Baker}, {Balaguer-N{\'u}{\~n}ez}, {Balm}, {Barache}, {Barata}, {Barbato},
  {Barblan}, {Barklem}, {Barrado}, {Barros}, {Barstow}, {Bartholom{\'e}
  Mu{\~n}oz}, {Bassilana}, {Becciani}, {Bellazzini}, {Berihuete}, {Bertone},
  {Bianchi}, {Bienaym{\'e}}, {Blanco-Cuaresma}, {Boch}, {Boeche}, {Bombrun},
  {Borrachero}, {Bossini}, {Bouquillon}, {Bourda}, {Bragaglia}, {Bramante},
  {Breddels}, {Bressan}, {Brouillet}, {Br{\"u}semeister}, {Brugaletta},
  {Bucciarelli}, {Burlacu}, {Busonero}, {Butkevich}, {Buzzi}, {Caffau},
  {Cancelliere}, {Cannizzaro}, {Cantat-Gaudin}, {Carballo}, {Carlucci},
  {Carrasco}, {Casamiquela}, {Castellani}, {Castro-Ginard}, {Charlot},
  {Chemin}, {Chiavassa}, {Cocozza}, {Costigan}, {Cowell}, {Crifo}, {Crosta},
  {Crowley}, {Cuypers}, {Dafonte}, {Damerdji}, {Dapergolas}, {David}, {David},
  {de Laverny}, {De Luise}, {De March}, {de Martino}, {de Souza}, {de Torres},
  {Debosscher}, {del Pozo}, {Delbo}, {Delgado}, {Delgado}, {Di Matteo},
  {Diakite}, {Diener}, {Distefano}, {Dolding}, {Drazinos}, {Dur{\'a}n},
  {Edvardsson}, {Enke}, {Eriksson}, {Esquej}, {Eynard Bontemps}, {Fabre},
  {Fabrizio}, {Faigler}, {Falc{\~a}o}, {Farr{\`a}s Casas}, {Federici},
  {Fedorets}, {Fernique}, {Figueras}, {Filippi}, {Findeisen}, {Fonti},
  {Fraile}, {Fraser}, {Fr{\'e}zouls}, {Gai}, {Galleti}, {Garabato},
  {Garc{\'\i}a-Sedano}, {Garofalo}, {Garralda}, {Gavel}, {Gavras}, {Gerssen},
  {Geyer}, {Giacobbe}, {Gilmore}, {Girona}, {Giuffrida}, {Glass}, {Gomes},
  {Granvik}, {Gueguen}, {Guerrier}, {Guiraud}, {Guti{\'e}rrez-S{\'a}nchez},
  {Haigron}, {Hatzidimitriou}, {Hauser}, {Haywood}, {Heiter}, {Helmi}, {Heu},
  {Hilger}, {Hobbs}, {Hofmann}, {Holland}, {Huckle}, {Hypki}, {Icardi},
  {Jan{\ss}en}, {Jevardat de Fombelle}, {Jonker}, {Juh{\'a}sz}, {Julbe},
  {Karampelas}, {Kewley}, {Klar}, {Kochoska}, {Kohley}, {Kolenberg},
  {Kontizas}, {Kontizas}, {Koposov}, {Kordopatis}, {Kostrzewa-Rutkowska},
  {Koubsky}, {Lambert}, {Lanza}, {Lasne}, {Lavigne}, {Le Fustec}, {Le
  Poncin-Lafitte}, {Lebreton}, {Leccia}, {Leclerc}, {Lecoeur-Taibi},
  {Lenhardt}, {Leroux}, {Liao}, {Licata}, {Lindstr{\o}m}, {Lister}, {Livanou},
  {Lobel}, {L{\'o}pez}, {Managau}, {Mann}, {Mantelet}, {Marchal}, {Marchant},
  {Marconi}, {Marinoni}, {Marschalk{\'o}}, {Marshall}, {Martino}, {Marton},
  {Mary}, {Massari}, {Matijevi{\v{c}}}, {Mazeh}, {McMillan}, {Messina},
  {Michalik}, {Millar}, {Molina}, {Molinaro}, {Moln{\'a}r}, {Montegriffo},
  {Mor}, {Morbidelli}, {Morel}, {Morris}, {Mulone}, {Muraveva}, {Musella},
  {Nelemans}, {Nicastro}, {Noval}, {O'Mullane}, {Ord{\'e}novic},
  {Ord{\'o}{\~n}ez-Blanco}, {Osborne}, {Pagani}, {Pagano}, {Pailler},
  {Palacin}, {Palaversa}, {Panahi}, {Pawlak}, {Piersimoni}, {Pineau}, {Plachy},
  {Plum}, {Poggio}, {Poujoulet}, {Pr{\v{s}}a}, {Pulone}, {Racero}, {Ragaini},
  {Rambaux}, {Ramos-Lerate}, {Regibo}, {Reyl{\'e}}, {Riclet}, {Ripepi}, {Riva},
  {Rivard}, {Rixon}, {Roegiers}, {Roelens}, {Romero-G{\'o}mez}, {Rowell},
  {Royer}, {Ruiz-Dern}, {Sadowski}, {Sagrist{\`a} Sell{\'e}s}, {Sahlmann},
  {Salgado}, {Salguero}, {Sanna}, {Santana-Ros}, {Sarasso}, {Savietto},
  {Schultheis}, {Sciacca}, {Segol}, {Segovia}, {S{\'e}gransan}, {Shih},
  {Siltala}, {Silva}, {Smart}, {Smith}, {Solano}, {Solitro}, {Sordo}, {Soria
  Nieto}, {Souchay}, {Spagna}, {Spoto}, {Stampa}, {Steele},
  {Steidelm{\"u}ller}, {Stephenson}, {Stoev}, {Suess}, {Surdej}, {Szabados},
  {Szegedi-Elek}, {Tapiador}, {Taris}, {Tauran}, {Taylor}, {Teixeira},
  {Terrett}, {Teyssand ier}, {Thuillot}, {Titarenko}, {Torra Clotet}, {Turon},
  {Ulla}, {Utrilla}, {Uzzi}, {Vaillant}, {Valentini}, {Valette}, {van Elteren},
  {Van Hemelryck}, {van Leeuwen}, {Vaschetto}, {Vecchiato}, {Veljanoski},
  {Viala}, {Vicente}, {Vogt}, {von Essen}, {Voss}, {Votruba}, {Voutsinas},
  {Walmsley}, {Weiler}, {Wertz}, {Wevers}, {Wyrzykowski}, {Yoldas},
  {{\v{Z}}erjal}, {Ziaeepour}, {Zorec}, {Zschocke}, {Zucker}, {Zurbach}, \&
  {Zwitter}}]{GAIA-DR2-2018}
{Gaia Collaboration}, {Brown}, A.~G.~A., {Vallenari}, A., {et~al.} 2018, \aap,
  616, A1

\bibitem[{{Gonzalez} {et~al.}(2015){Gonzalez}, {Laibe}, {Maddison}, {Pinte}, \&
  {M{\'e}nard}}]{Gonzalez2015}
{Gonzalez}, J.~F., {Laibe}, G., {Maddison}, S.~T., {Pinte}, C., \&
  {M{\'e}nard}, F. 2015, \mnras, 454, L36

\bibitem[{{Gravity Collaboration} {et~al.}(2017){Gravity Collaboration},
  {Abuter}, {Accardo}, {Amorim}, {Anugu}, {{\'A}vila}, {Azouaoui}, {Benisty},
  {Berger}, {Blind}, {Bonnet}, {Bourget}, {Brandner}, {Brast}, {Buron},
  {Burtscher}, {Cassaing}, {Chapron}, {Choquet}, {Cl{\'e}net}, {Collin},
  {Coud{\'e} Du Foresto}, {de Wit}, {de Zeeuw}, {Deen},
  {Delplancke-Str{\"o}bele}, {Dembet}, {Derie}, {Dexter}, {Duvert}, {Ebert},
  {Eckart}, {Eisenhauer}, {Esselborn}, {F{\'e}dou}, {Finger}, {Garcia}, {Garcia
  Dabo}, {Garcia Lopez}, {Gendron}, {Genzel}, {Gillessen}, {Gonte}, {Gordo},
  {Grould}, {Gr{\"o}zinger}, {Guieu}, {Haguenauer}, {Hans}, {Haubois}, {Haug},
  {Haussmann}, {Henning}, {Hippler}, {Horrobin}, {Huber}, {Hubert}, {Hubin},
  {Hummel}, {Jakob}, {Janssen}, {Jochum}, {Jocou}, {Kaufer}, {Kellner},
  {Kendrew}, {Kern}, {Kervella}, {Kiekebusch}, {Klein}, {Kok}, {Kolb}, {Kulas},
  {Lacour}, {Lapeyr{\`e}re}, {Lazareff}, {Le Bouquin}, {L{\`e}na}, {Lenzen},
  {L{\'e}v{\^e}que}, {Lippa}, {Magnard}, {Mehrgan}, {Mellein}, {M{\'e}rand},
  {Moreno-Ventas}, {Moulin}, {M{\"u}ller}, {M{\"u}ller}, {Neumann}, {Oberti},
  {Ott}, {Pallanca}, {Panduro}, {Pasquini}, {Paumard}, {Percheron}, {Perraut},
  {Perrin}, {Pfl{\"u}ger}, {Pfuhl}, {Phan Duc}, {Plewa}, {Popovic}, {Rabien},
  {Ram{\'\i}rez}, {Ramos}, {Rau}, {Riquelme}, {Rohloff}, {Rousset},
  {Sanchez-Bermudez}, {Scheithauer}, {Sch{\"o}ller}, {Schuhler}, {Spyromilio},
  {Straubmeier}, {Sturm}, {Suarez}, {Tristram}, {Ventura}, {Vincent},
  {Waisberg}, {Wank}, {Weber}, {Wieprecht}, {Wiest}, {Wiezorrek}, {Wittkowski},
  {Woillez}, {Wolff}, {Yazici}, {Ziegler}, \& {Zins}}]{GRAVITY2017}
{Gravity Collaboration}, {Abuter}, R., {Accardo}, M., {et~al.} 2017, \aap, 602,
  A94

\bibitem[{{Hauschildt} {et~al.}(1999){Hauschildt}, {Allard}, \&
  {Baron}}]{Hauschildt1999}
{Hauschildt}, P.~H., {Allard}, F., \& {Baron}, E. 1999, \apj, 512, 377

\bibitem[{{Johansen} {et~al.}(2014){Johansen}, {Blum}, {Tanaka}, {Ormel},
  {Bizzarro}, \& {Rickman}}]{Johansen2014}
{Johansen}, A., {Blum}, J., {Tanaka}, H., {et~al.} 2014, in Protostars and
  Planets VI, ed. H.~{Beuther}, R.~S. {Klessen}, C.~P. {Dullemond}, \&
  T.~{Henning}, 547

\bibitem[{{Lacour} {et~al.}(2019){Lacour}, {Dembet}, {Abuter}, {F{\'e}dou},
  {Perrin}, {Choquet}, {Pfuhl}, {Eisenhauer}, {Woillez}, {Cassaing},
  {Wieprecht}, {Ott}, {Wiezorrek}, {Tristram}, {Wolff}, {Ram{\'\i}rez},
  {Haubois}, {Perraut}, {Straubmeier}, {Brand ner}, \& {Amorim}}]{Lacour2019}
{Lacour}, S., {Dembet}, R., {Abuter}, R., {et~al.} 2019, \aap, 624, A99

\bibitem[{{Lambrechts} \& {Johansen}(2012)}]{Lambrechts2012}
{Lambrechts}, M. \& {Johansen}, A. 2012, \aap, 544, A32

\bibitem[{{Langlois} {et~al.}(2018){Langlois}, {Pohl}, {Lagrange}, {Maire},
  {Mesa}, {Boccaletti}, {Gratton}, {Denneulin}, {Klahr}, {Vigan}, {Benisty},
  {Dominik}, {Bonnefoy}, {Menard}, {Avenhaus}, {Cheetham}, {Van Boekel}, {de
  Boer}, {Chauvin}, {Desidera}, {Feldt}, {Galicher}, {Ginski}, {Girard},
  {Henning}, {Janson}, {Kopytova}, {Kral}, {Ligi}, {Messina}, {Peretti},
  {Pinte}, {Sissa}, {Stolker}, {Zurlo}, {Magnard}, {Blanchard}, {Buey},
  {Suarez}, {Cascone}, {Moller-Nilsson}, {Weber}, {Petit}, \&
  {Pragt}}]{Langlois2018}
{Langlois}, M., {Pohl}, A., {Lagrange}, A.~M., {et~al.} 2018, \aap, 614, A88

\bibitem[{{Lapeyrere} {et~al.}(2014){Lapeyrere}, {Kervella}, {Lacour},
  {Azouaoui}, {Garcia-Dabo}, {Perrin}, {Eisenhauer}, {Perraut}, {Straubmeier},
  {Amorim}, \& {Brandner}}]{Lapeyrere2014}
{Lapeyrere}, V., {Kervella}, P., {Lacour}, S., {et~al.} 2014, Society of
  Photo-Optical Instrumentation Engineers (SPIE) Conference Series, Vol. 9146,
  {GRAVITY data reduction software}, 91462D

\bibitem[{{Lazareff} {et~al.}(2017){Lazareff}, {Berger}, {Kluska}, {Le
  Bouquin}, {Benisty}, {Malbet}, {Koen}, {Pinte}, {Thi}, {Absil}, {Baron},
  {Delboulb{\'e}}, {Duvert}, {Isella}, {Jocou}, {Juhasz}, {Kraus}, {Lachaume},
  {M{\'e}nard}, {Millan-Gabet}, {Monnier}, {Moulin}, {Perraut}, {Rochat},
  {Soulez}, {Tallon}, {Thi{\'e}baut}, {Traub}, \& {Zins}}]{Lazareff2017}
{Lazareff}, B., {Berger}, J.~P., {Kluska}, J., {et~al.} 2017, \aap, 599, A85

\bibitem[{{Le Bouquin} {et~al.}(2011){Le Bouquin}, {Berger}, {Lazareff},
  {Zins}, {Haguenauer}, {Jocou}, {Kern}, {Millan-Gabet}, {Traub}, {Absil},
  {Augereau}, {Benisty}, {Blind}, {Bonfils}, {Bourget}, {Delboulbe},
  {Feautrier}, {Germain}, {Gitton}, {Gillier}, {Kiekebusch}, {Kluska},
  {Knudstrup}, {Labeye}, {Lizon}, {Monin}, {Magnard}, {Malbet}, {Maurel},
  {M{\'e}nard}, {Micallef}, {Michaud}, {Montagnier}, {Morel}, {Moulin},
  {Perraut}, {Popovic}, {Rabou}, {Rochat}, {Rojas}, {Roussel}, {Roux},
  {Stadler}, {Stefl}, {Tatulli}, \& {Ventura}}]{Lebouquin2011}
{Le Bouquin}, J.-B., {Berger}, J.-P., {Lazareff}, B., {et~al.} 2011, \aap, 535,
  A67

\bibitem[{{Long} {et~al.}(2018){Long}, {Pinilla}, {Herczeg}, {Harsono},
  {Dipierro}, {Pascucci}, {Hendler}, {Tazzari}, {Ragusa}, {Salyk}, {Edwards},
  {Lodato}, {van de Plas}, {Johnstone}, {Liu}, {Boehler}, {Cabrit}, {Manara},
  {Menard}, {Mulders}, {Nisini}, {Fischer}, {Rigliaco}, {Banzatti}, {Avenhaus},
  \& {Gully-Santiago}}]{Long2018}
{Long}, F., {Pinilla}, P., {Herczeg}, G.~J., {et~al.} 2018, \apj, 869, 17

\bibitem[{{Lopez} {et~al.}(2018){Lopez}, {Lagarde}, {Matter}, {Agocs},
  {Allouche}, {Antonelli}, {Augereau}, {Bailet}, {Berio}, {Bettonvil},
  {Beckmann}, {van Boekel}, {Bresson}, {Bristow}, {Cruzalebes}, {Delbo},
  {Dominik}, {Elswijk}, {Fantei}, {Glindemann}, {Heininger}, {Hofmann},
  {Hogerheijde}, {Hron}, {Jaffe}, {Kroes}, {Laun}, {Lehmitz}, {Meilland},
  {Meisenheimer}, {Millour}, {Morel}, {Neumann}, {Pantin}, {Petrov},
  {Robbe-Dubois}, {Schertl}, {Schoeller}, {Wolf}, {Zins}, {Henning}, {Stee}, \&
  {Weigelt}}]{Lopez2018}
{Lopez}, B., {Lagarde}, S., {Matter}, A., {et~al.} 2018, in Society of
  Photo-Optical Instrumentation Engineers (SPIE) Conference Series, Vol. 10701,
  \procspie, 107010Z

\bibitem[{{Lor{\'e}n-Aguilar} \& {Bate}(2015)}]{Loren-Aguilar2015}
{Lor{\'e}n-Aguilar}, P. \& {Bate}, M.~R. 2015, \mnras, 453, L78

\bibitem[{{Manset} {et~al.}(2009){Manset}, {Bastien}, {M{\'e}nard}, {Bertout},
  {Le van Suu}, \& {Boivin}}]{Manset2009}
{Manset}, N., {Bastien}, P., {M{\'e}nard}, F., {et~al.} 2009, \aap, 499, 137

\bibitem[{{Marino} {et~al.}(2015){Marino}, {Perez}, \& {Casassus}}]{Marino2015}
{Marino}, S., {Perez}, S., \& {Casassus}, S. 2015, \apjl, 798, L44

\bibitem[{{Morel} \& {Magnenat}(1978)}]{morel1978}
{Morel}, M. \& {Magnenat}, P. 1978, \aaps, 34, 477

\bibitem[{{Ormel} \& {Klahr}(2010)}]{Ormel2010}
{Ormel}, C.~W. \& {Klahr}, H.~H. 2010, \aap, 520, A43

\bibitem[{{Pani{\'c}} {et~al.}(2009){Pani{\'c}}, {Hogerheijde}, {Wilner}, \&
  {Qi}}]{Panic2009}
{Pani{\'c}}, O., {Hogerheijde}, M.~R., {Wilner}, D., \& {Qi}, C. 2009, \aap,
  501, 269

\bibitem[{{Pinte} {et~al.}(2016){Pinte}, {Dent}, {M{\'e}nard}, {Hales}, {Hill},
  {Cortes}, \& {de Gregorio-Monsalvo}}]{Pinte2016}
{Pinte}, C., {Dent}, W.~R.~F., {M{\'e}nard}, F., {et~al.} 2016, \apj, 816, 25

\bibitem[{{Pinte} {et~al.}(2009){Pinte}, {Harries}, {Min}, {Watson},
  {Dullemond}, {Woitke}, {M{\'e}nard}, \& {Dur{\'a}n-Rojas}}]{Pinte2009}
{Pinte}, C., {Harries}, T.~J., {Min}, M., {et~al.} 2009, \aap, 498, 967

\bibitem[{{Pinte} {et~al.}(2008{\natexlab{a}}){Pinte}, {M{\'e}nard}, {Berger},
  {Benisty}, \& {Malbet}}]{Pinte2008b}
{Pinte}, C., {M{\'e}nard}, F., {Berger}, J.~P., {Benisty}, M., \& {Malbet}, F.
  2008{\natexlab{a}}, \apjl, 673, L63

\bibitem[{{Pinte} {et~al.}(2006){Pinte}, {M{\'e}nard}, {Duch{\^e}ne}, \&
  {Bastien}}]{pinte2006}
{Pinte}, C., {M{\'e}nard}, F., {Duch{\^e}ne}, G., \& {Bastien}, P. 2006, \aap,
  459, 797

\bibitem[{{Pinte} {et~al.}(2008{\natexlab{b}}){Pinte}, {Padgett}, {M{\'e}nard},
  {Stapelfeldt}, {Schneider}, {Olofsson}, {Pani{\'c}}, {Augereau},
  {Duch{\^e}ne}, {Krist}, {Pontoppidan}, {Perrin}, {Grady}, {Kessler-Silacci},
  {van Dishoeck}, {Lommen}, {Silverstone}, {Hines}, {Wolf}, {Blake}, {Henning},
  \& {Stecklum}}]{Pinte2008}
{Pinte}, C., {Padgett}, D.~L., {M{\'e}nard}, F., {et~al.} 2008{\natexlab{b}},
  \aap, 489, 633

\bibitem[{{Pohl} {et~al.}(2017){Pohl}, {Benisty}, {Pinilla}, {Ginski}, {de
  Boer}, {Avenhaus}, {Henning}, {Zurlo}, {Boccaletti}, {Augereau}, {Birnstiel},
  {Dominik}, {Facchini}, {Fedele}, {Janson}, {Keppler}, {Kral}, {Langlois},
  {Ligi}, {Maire}, {M{\'e}nard}, {Meyer}, {Pinte}, {Quanz}, {Sauvage},
  {Sezestre}, {Stolker}, {Szul{\'a}gyi}, {van Boekel}, {van der Plas},
  {Villenave}, {Baruffolo}, {Baudoz}, {Le Mignant}, {Maurel}, {Ramos}, \&
  {Weber}}]{Pohl2017}
{Pohl}, A., {Benisty}, M., {Pinilla}, P., {et~al.} 2017, \apj, 850, 52

\bibitem[{{Pollack} {et~al.}(1996){Pollack}, {Hubickyj}, {Bodenheimer},
  {Lissauer}, {Podolak}, \& {Greenzweig}}]{Pollack1996}
{Pollack}, J.~B., {Hubickyj}, O., {Bodenheimer}, P., {et~al.} 1996, \icarus,
  124, 62

\bibitem[{{Rice} \& {Armitage}(2003)}]{Rice2003}
{Rice}, W.~K.~M. \& {Armitage}, P.~J. 2003, \apjl, 598, L55

\bibitem[{{Siess} {et~al.}(2000){Siess}, {Dufour}, \& {Forestini}}]{Siess2000}
{Siess}, L., {Dufour}, E., \& {Forestini}, M. 2000, \aap, 358, 593

\bibitem[{{van der Marel} {et~al.}(2018){van der Marel}, {Williams}, {Ansdell},
  {Manara}, {Miotello}, {Tazzari}, {Testi}, {Hogerheijde}, {Bruderer}, {van
  Terwisga}, \& {van Dishoeck}}]{vanderMarel2018}
{van der Marel}, N., {Williams}, J.~P., {Ansdell}, M., {et~al.} 2018, \apj,
  854, 177

\bibitem[{{Vural} {et~al.}(2012){Vural}, {Kreplin}, {Kraus}, {Weigelt},
  {Driebe}, {Benisty}, {Dugu{\'e}}, {Massi}, {Monin}, \& {Vannier}}]{Vural2012}
{Vural}, J., {Kreplin}, A., {Kraus}, S., {et~al.} 2012, \aap, 543, A162

\bibitem[{{Woillez} {et~al.}(2019){Woillez}, {Abad}, {Abuter}, {Aller
  Carpentier}, {Alonso}, {Andolfato}, {Barriga}, {Berger}, {Beuzit}, {Bonnet},
  {Bourdarot}, {Bourget}, {Brast}, {Caniguante}, {Cottalorda}, {Darr{\'e}},
  {Delabre}, {Delboulb{\'e}}, {Delplancke-Str{\"o}bele}, {Dembet}, {Donaldson},
  {Dorn}, {Dupeyron}, {Dupuy}, {Egner}, {Eisenhauer}, {Fischer}, {Frank},
  {Fuenteseca}, {Gitton}, {Gont{\'e}}, {Guerlet}, {Guieu}, {Gutierrez},
  {Haguenauer}, {Haimerl}, {Haubois}, {Heritier}, {Huber}, {Hubin}, {Jolley},
  {Jocou}, {Kirchbauer}, {Kolb}, {Kosmalski}, {Krempl}, {Le Bouquin}, {Le
  Louarn}, {Lilley}, {Lopez}, {Magnard}, {Mclay}, {Meilland}, {Meister},
  {Merand}, {Moulin}, {Pasquini}, {Paufique}, {Percheron}, {Pettazzi}, {Pfuhl},
  {Phan}, {Pirani}, {Quentin}, {Rakich}, {Ridings}, {Riedel}, {Reyes},
  {Rochat}, {Santos Tom{\'a}s}, {Schmid}, {Schuhler}, {Shchekaturov}, {Seidel},
  {Soenke}, {Stadler}, {Stephan}, {Su{\'a}rez}, {Todorovic}, {Valdes},
  {Verinaud}, {Zins}, \& {Z{\'u}{\~n}iga-Fern{\'a}ndez}}]{Woillez2019}
{Woillez}, J., {Abad}, J.~A., {Abuter}, R., {et~al.} 2019, \aap, 629, A41

\bibitem[{{Yen} {et~al.}(2018){Yen}, {Koch}, {Manara}, {Miotello}, \&
  {Testi}}]{Yen2018}
{Yen}, H.-W., {Koch}, P.~M., {Manara}, C.~F., {Miotello}, A., \& {Testi}, L.
  2018, \aap, 616, A100

\bibitem[{{Youdin} \& {Goodman}(2005)}]{Youdin2005}
{Youdin}, A.~N. \& {Goodman}, J. 2005, \apj, 620, 459

\bibitem[{{Zhang} {et~al.}(2015){Zhang}, {Blake}, \& {Bergin}}]{Zhang2015}
{Zhang}, K., {Blake}, G.~A., \& {Bergin}, E.~A. 2015, \apjl, 806, L7

\end{thebibliography}

\begin{appendix}
\section{The M70 model}
\begin{figure*}[t!]%
\centering
\includegraphics[width=9cm]{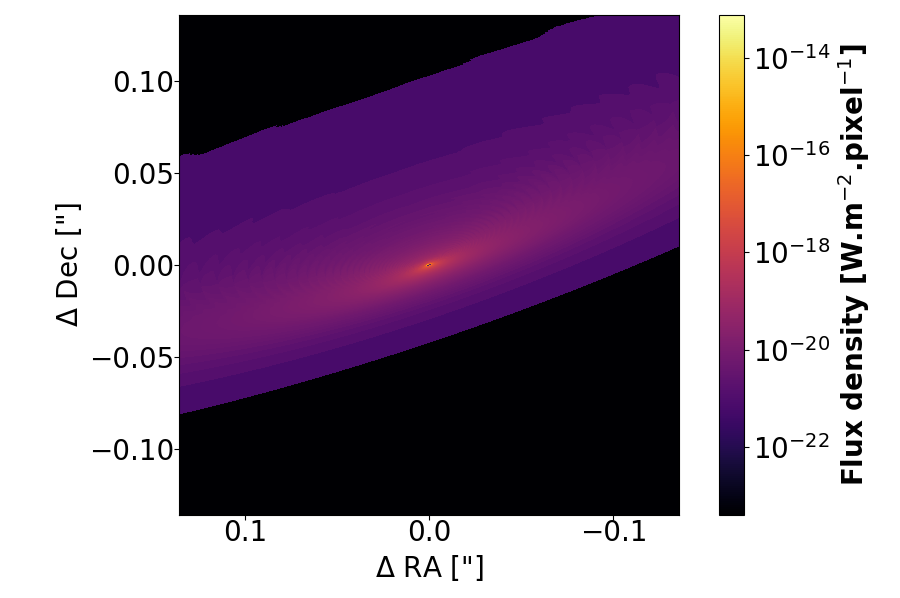}
\includegraphics[width=9cm]{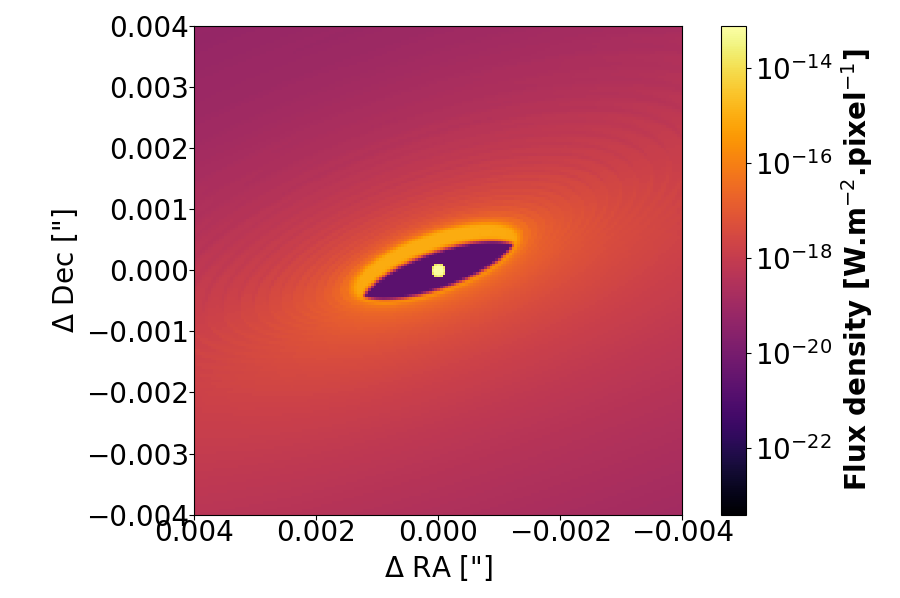}
\caption{Intensity maps produced by MCFOST for our model M70 of RY~Lup in the K band for the whole disk (left) and  a zoom into the central region  at the milliarcsecond
scale (right). }
% \label{fig:M70}
\end{figure*}
\end{appendix}

\end{document}